\documentclass[aps,amsmath,amssymb,twocolumn,showpacs]{revtex4}
\usepackage{graphicx}
\usepackage{bm}

\newif\ifdraft
\drafttrue              

    \usepackage{amssymb,amsmath,amsfonts}
    \usepackage{graphicx}
    \usepackage{dsfont}
    \usepackage{mathrsfs}
    \usepackage{natbib}       
    \usepackage[dvips]{color} 
\ifdraft
   \newcommand{\JRE}[1]{$\footnotemark\footnotetext{JRE: #1}$}
   \newcommand{\PC}[1]{$\footnotemark\footnotetext{PC: #1}$}

   \newcommand{\fix}[1] {{\color{magenta} #1 }}
   \newcommand{\file}[1]{$\footnotemark\footnotetext{{ file:} #1}$}
   \newcommand{\PublicPrivate}[2]
        {{\color{blue}#2}}%
\else
   \newcommand{\JRE}[1]{}
   \newcommand{\PC}[1]{}

   \newcommand{\fix}[1]{}
   \newcommand{\file}[1]{}
   \newcommand{\PublicPrivate}[2]{#1}
\fi

\newcommand{\refeq}   [1] {(\ref{#1})}

\newcommand{\reffig}  [1] {Fig.~\ref{#1}}
\newcommand{\reffigs} [2] {Figs.~\ref{#1} and~\ref{#2}}

\newcommand{\refsect} [1] {sect.~\ref{#1}}

\newcommand{\beq}{\begin{equation}}
\newcommand{\eeq}{\end{equation}}
\newcommand{\ee}[1] {\label{#1} \end{equation}}
\newcommand{\bea}{\begin{eqnarray}}

\newcommand{\eea}{\end{eqnarray}}

\newcommand{\VectorII}[2]{
   \begin{pmatrix}
             {#1} \cr
             {#2} \cr
   \end{pmatrix}
                         }

\renewcommand{\det}{\mbox{\rm det}\,}

\newcommand{\tr}{\mbox{\rm tr}\,}

\begin{document}

\title{Fluctuations in classical sum rules}
\author{John R.~Elton$^{1,2}$, Arul Lakshminarayan$^{1,3}$, and Steven Tomsovic$^{1,3,4}$} 
\affiliation{$^1$ Max-Planck-Institut f\"ur Physik komplexer Systeme, D-01187 Dresden, Germany}
\affiliation{$^2$ Department of Physics, Cornell University, Ithaca, NY 14583 USA (current address)}
\affiliation{$^3$ Department of Physics, Indian Institute of Technology Madras, Chennai, 600 036,
India (permanent address)}    
\affiliation{$^4$ Department of Physics and Astronomy; Washington State University; Pullman, WA 99164-2814 (permanent address)} 
\date{\today}

\begin{abstract}
Classical sum rules arise in a wide variety of physical contexts.  Asymptotic expressions have been derived for many of these sum rules in the limit of long orbital period (or large action).  Although sum rule convergence may well be exponentially rapid for chaotic systems in a global sense with time, individual contributions to the sums may fluctuate with a width which diverges in time.  Our interest is in the global convergence of sum rules as well as their local fluctuations.  It turns out that a simple version of a lazy baker map gives an ideal system in which classical sum rules, their corrections, and their fluctuations can be worked out analytically.  This is worked out in detail for the Hannay-Ozorio sum rule. In this particular case the rate of convergence of the sum rule is found to be governed by the Pollicott-Ruelle resonances, and both local and global boundaries for which the sum rule may converge are given.  In addition, the width of the fluctuations is considered and worked out analytically, and it is shown to have an interesting dependence on the location of the region over which the sum rule is applied. It is also found that as the region of application is decreased in size the fluctuations grow.  This suggests a way of controlling the length scale of the fluctuations by considering a time dependent phase space volume, which for the lazy baker map decreases exponentially rapidly with time.  

\end{abstract}

\pacs{05.45.-a, 05.45.Ac,45.05.+x,03.65.Sq} \maketitle

\section{Introduction}
\label{sec:introduction}

Classical sum rules play an important role in a number of physical contexts.  It is interesting to have a case of a non-uniformly spreading chaotic system where their limits, fluctuations, and convergences can be studied analytically, which is given in this paper.  We focus on the sum rule of Hannay and Ozorio de Almeida, who were motivated by a desire to understand the two-point quantum density of states correlator~\cite{Hannay84}.   Their derivation relied on the Principle of Uniformity, which states that the periodic orbits, weighted naturally are uniformly dense in phase space.   There are many others; see, for example, the sum rules in Refs.~\cite{Argaman96,Sieber99,Richter02} which involve return probability, the connection of two points in coordinate space, and fixed orientations of initial and final velocities, respectively.  They arise in mesoscopic conductance, diffraction contributions to spectral fluctuations, and chaotic quantum transport, also respectively. 

In one possible form of the Hannay-Ozorio sum rule~\cite{OzorioBook} for chaotic systems, it is the inverses of the stability matrix determinants for the periodic orbits of period $\tau$, $\left|{\rm Det}\left( {\mathbf M}_\tau - {\mathbf 1} \right)  \right|^{-1}$, which are summed.  For completely chaotic systems the finite-time stability exponents, which roughly determine the value of each determinant, converge as $\tau\rightarrow \infty$ to a single set. Nevertheless, the determinant's fluctuations from one periodic orbit to the next grow without bound in the same limit.   This feature is made even more curious by the expectation that the Hannay-Ozorio sum rule converges exponentially rapidly.  See a preprint by Pollicott~\cite{Pollicott01} for a theorem regarding the sum rule's convergence.

It is known that smooth classical functions have exponential decays in fully chaotic systems toward their ergodic averages, which are governed by the Pollicott-Ruelle (PR) resonances~\cite{Pollicott85,Pollicott86,Ruelle86,Ruelle87} associated with the Perron-Frobenius operator.  It is natural to ask whether the kinds of classical sum rules encountered always converge to their limiting values with corrections that decay exponentially according to the leading PR resonances.  The work of Andreev et al.~\cite{Andreev96} would suggest that as long as there is a gap in the spectra of the Perron-Frobenius operators, there is no other possibility.  

A first step is taken here in approaching this general line of questioning by considering a form of the Hannay-Ozorio sum rule for fixed period in maps.  Its convergence rate and local fluctuations are considered in detail in a simple version of a lazy baker map introduced by  Balazs and one of us (AL)~\cite{Lakshminarayan93,Lakshminarayan93a}.  Although the usual baker map is especially simple, it is also non-generic in that its lacks the essential fluctuations of interest, whereas the lazy baker map possesses fluctuations and still turns out to be analytically tractable.  The fluctuations of the inverse determinant were studied briefly in Ref.~\cite{Tomsovic07}, where they were demonstrated, not surprisingly, to be extremely sensitive to even tiny islands of stability.  In the form of the lazy baker map studied here, all orbits, with one exception of  a marginal orbit of period two, are unstable, and there is no ambiguity concerning whether the system is completely chaotic. 

The organization of this paper is as follows: in Sect.~\ref{sec:notations} notations are specified and definitions of various quantities of interest are given.  Section \ref{sec:srmap} introduces a simple version of a lazy baker map (named the SR map), which is studied in detail, and gives basic counting results for fixed points and a method of subdividing the phase space into local regions.  This is followed in Sect.~\ref{sec:statistics} by a derivation of the main results for the local fluctuation and convergence properties of the Hannay-Ozorio sum rule.

\section{Measurements of chaotic systems}
\label{sec:notations}

Dynamical systems theory has generated a number of ways to specify the complexity of a chaotic mapping.  Three of the more familiar concepts to physicists are the topological entropy, $h_T$, the metric or Kolmogorov-Sinai (KS) entropy, and the Lyapunov exponent, $\lambda_L$~\cite{Jacobs98,Ott02}.  The topological entropy is designed to measure the information content of the optimal partition of the dynamics.  It turns out for that a class of systems known as Axiom A, the limiting value of the exponential rate of increase in the number of fixed points with iteration number gives this entropy.  The KS entropy can be thought of in a similar way, except that it is weighted.   And finally, the Lyapunov exponent measures the exponential separation of neighboring initial conditions.

For the purpose of studying the fluctuations of classical sum rules, the main quantities of interest tend to be the number of fixed points, their finite-time stability exponents, and their probability densities and moments, all of which can be considered in both a global and local phase space context.  We do not worry as to what exact relations exist between each of these measures for a given system, but they are closely related where they do not have an identical counterpart and it is useful to relate our results to some of these quantities when they are known.

\subsection{Basic quantities}
\label{sec:notationsa}

The notation $N_{\tau}$ denotes the number of fixed points at integer time $\tau$ taken over the full phase space.  This is distinguished from a local count of fixed points by writing $N_{\tau}(s,k)$ where the parameters $s$ and $k$ conveniently specify the location and size respectively of the local phase space volume in question for the SR map. Additionally, of great importance in this work are the sums over fixed points appearing in one form of the Hannay-Ozorio sum rule, which are given the notation
\begin{eqnarray}
F_{\tau} &=& \sum_{f.p.} \frac{1}{\left|{\rm Det}({\bf M}_\tau-{\bf 1})\right|}\qquad\ \ \ ({\rm global})\nonumber \\
F_{\tau}(s,k) &=& \sum_{f.p.\in (s,k)} \frac{1}{\left|{\rm Det}({\bf M}_\tau-{\bf 1})\right|}\ \ \ ({\rm local})
\label{HannayOzorio}
\end{eqnarray}
where ``f.p.'' denotes fixed points in the specified region of phase space and
\begin{equation}
\label{stabilitymatrix}
{\bf M}_\tau(q_0,p_0) = \prod_{i=0}^{\tau-1}{{\bf M}(q_i,p_i)}
\end{equation}
is the Jacobian stability matrix along the trajectory fixed by the set of iterates $\{ q_i,p_i \}$ of the initial conditions $(q_0,p_0)$; the notation for initial conditions is mostly left suppressed.  A related quantity which is sometimes of interest is the finite-time stability exponent, given here for a map with single position and single momentum coordinates
\begin{eqnarray}
\label{deflambda}
\lambda(q_0,p_0;\tau) &=& {1\over \tau} \ln \left ({\left|\mbox{\tr}[{\bf M}_\tau]\right| + \sqrt{\mbox{\tr}[{\bf M}_\tau]^2-4}\over 2}\right) \nonumber \\
& \approx& {1\over \tau} \ln \left| \tr[{\bf M}_\tau] \right| \approx \frac{1}{\tau}\ln {\left|{\rm Det}({\bf M}_\tau-{\bf 1})\right|} \nonumber \\
\end{eqnarray}
where $\tr(...)$ denotes the trace operation.  For long times, the approximate relations for $\lambda$ tend exponentially quickly to the first relation.

\subsection{Convergence and measuring sum rule fluctuations}
\label{sec:fluctuations}

The Hannay-Ozorio sum rule in this context and notation (also referred to as the uniformity principle) reads $F_{\tau}(s,k) \rightarrow {\cal V}_k$ where ${\cal V}_k$ is the phase space volume over which the fixed points are summed.  The absence of an $s$ dependence is the absence of a dependence on the location in phase space.  This result, which holds in both the local and global cases, emerges in the limit of long times.  The simplest determination of its convergence amounts to calculating the leading corrections to the sum as a function of $\{\tau,s,k\}$.  The expectation is that it should decrease exponentially with $\tau$ if $\{s,k\}$ are held fixed, although it is not obvious to us, a priori, at precisely which rate and with what length oscillations.  Ahead, it turns out to be governed by the leading Pollicott-Ruelle resonance, which interestingly enough, has a real part equal to the Lyapunov exponent in the SR map.

\subsubsection{Local convergence boundary}
\label{sect:averages}

It is also possible to consider the convergence with increasingly smaller local regions in the phase space.  By controlling the size of a region, the number of terms contributing to the local sum can be tuned for a given $\tau$.  Given that the individual stability determinants fluctuate with ever increasing width as $\tau$ increases, the question becomes, ``at which time on average do the corrections become smaller than the sum rule expectation, i.e.~the local phase space volume?''  As the regions shrink in size this time extends later, thus making it possible to find a shrinking volume as a function of $\tau$ that offsets the exponentially decaying corrections.  One can think of this ``${\cal V}(\tau)$'' as the boundary for the local application of the sum rule to be converged.  Both the global and the local boundary of convergence is given ahead.

\subsubsection{Moments}
\label{moments}

For the subregions of phase space, subtracting the local volume from the sum itself, call it $\tilde F_{\tau}(s,k)$, gives the leading corrections and fluctuating components of the sum rule.  A probability density for the values it takes on at fixed time over all regions can be defined, $P_{\tilde F_{\tau}}(x)$, which carries all information about convergence, local or global, and fluctuations.  In particular, our focus in this paper is on central moments of the density.  A very important case is the mean square deviation
\begin{equation}
\sigma^2(\tilde F_{\tau},k) = \langle \tilde F_{\tau}(s,k)^2 \rangle_s = \int {\rm d} x\ \hat x^2 P_{\tilde F_{\tau}}(x)
\end{equation} 
where $\hat x = x- \bar x$ and for which an asymptotic expression is derived in the case of the SR map.  More generally, the $n^{th}$ central moment is
\begin{equation}
{\cal M}_n(\tilde F_{\tau},k) = \langle \tilde F_{\tau}(s,k)^n \rangle_s = \int {\rm d}x\ \hat x^n P_{\tilde F_{\tau}}(x).
\end{equation}
The moments
\begin{equation}
{\cal M}_n( e^{\lambda\tau},k) = \langle e^{n\lambda(s,k;\tau)\tau} \rangle_s = \int {\rm d}x\ x^n P_{\exp(\lambda\tau)}(x)
\label{moment} 
\end{equation} 
which by Eq.~\refeq{deflambda} is associated with the probability density of the finite-time stability determinants, are distinctly different from the moments of the sum rule fluctuations.  Both cases are treated in this paper. However, as also shown, the probability density $P_{\tilde F_{\tau}}(x)$ asymptotically tends to a Gaussian density and only the first two moments (cumulants) are considered in detail.

\section{The Lazy Baker SR map}
\label{sec:srmap}

\begin{figure}[!ht]
 \includegraphics[width=.4\textwidth]{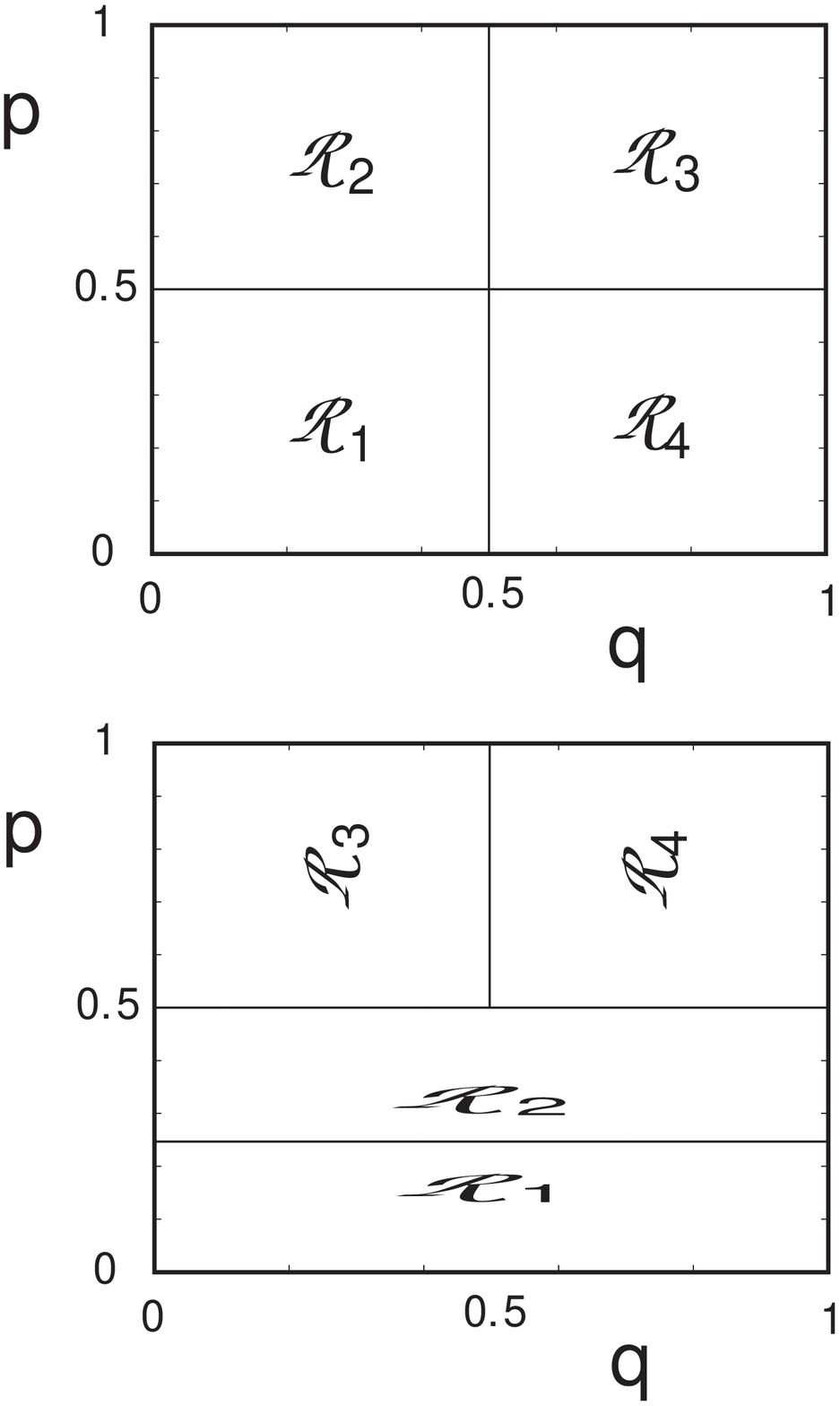}
  \caption{Partition of the unit square into four subregions, $\mathcal R_1$ through $\mathcal R_4$ (upper panel) and where each subregion maps after one iteration (lower panel).  As shown, $\mathcal R_1$ and $\mathcal R_2$ are squeezed and stretched whereas $\mathcal R_3$ and $\mathcal R_4$ are rotated.}
  \label{fig:regions}
 \end{figure}

 The Hannay-Ozorio sum rule fluctuations may be worked out exactly for the case of
  a simple dynamical system which is a modification of the usual baker's map.
 Lazy baker maps have previously been introduced~\cite{Lakshminarayan93} as a class of 2D
area-preserving maps. We study here a particular case called the SR map (stretch-rotate) which is chaotic over the whole measure and is defined on
the unit square in the usual position, momentum coordinates as follows: \begin{align} &\left.
\begin{array}{l}
q' \hspace{2 mm}= \hspace{2 mm} 2q \\
p' \hspace{2 mm} = \hspace{2 mm} p/2 \\
\end{array}  \right \} \hspace{5 mm} \textrm{if}\hspace{5 mm} 0 \leq q \leq 1/2, \nonumber \\
  &\left. \begin{array}{l}
q' \hspace{2 mm} = \hspace{2 mm} 1-p \\
p' \hspace{2 mm} = \hspace{2 mm} q \\
\end{array}  \right\} \hspace{5 mm} \textrm{if}\hspace{5 mm} 1/2 < q \leq
1. \end{align} The action of the map can be pictured most easily by
splitting the unit square into four equal subsquares, $\mathcal R_1$ through $\mathcal R_4$, shown
in \reffig{fig:regions}.
The region $\mathcal R_4$ is rotated uniquely into $ \mathcal R_3$ on the next
iteration and region $\mathcal R_3$ is rotated into $\mathcal R_2$. On the left half of the square,
points in $\mathcal R_2$ and $\mathcal R_1$ are compressed by a factor two along the $p$ axis and stretched by the same factor along the $q$ axis.

The combination of rotation and stretching in the same dynamical system gives rise to the possibility
of nonuniform hyperbolic motion or even non-hyperbolic motion.  For the SR map defined as above with the vertical cut in the middle of the square at $q = 0.5$, the motion is of the former
 kind.  As the cut is moved to the right it ceases to
be completely hyperbolic at the golden mean. In this paper we will only study the 
case when there are equal regions stretching and rotating, which is arguably the simplest ``exactly solvable" model of nonuniform hyperbolicity in an area preserving map. It admits a Markov partition of phase space and the dynamics is one of subshift of finite type on {\em three} symbols as shown in~\cite{Lakshminarayan93a}.  The atoms of the partition are $A=\mathcal{ R}_1 \bigcup \mathcal{ R}_2, \, B=\mathcal{ R}_3,$ and $C= \mathcal{ R}_4$.

The transition matrix is 
\begin{equation}
T_0= \left( \begin{array}{ccc}1&0&1\\1&0&0\\0&1&0 \end{array} \right)
\label{transition.matrix}
\end{equation}
whose $ij$ element, $t_{ij}$, is $1$ if there is a transition from $i$ to $j$
and $0$ otherwise. Here $i,j\, \in \ \{A,B,C\}$.
This has only topological information. Let $p_{ij}$ be the fraction of atom $i$ in atom $j$, on
one evolution of the map. This gives the transition probabilities for the 3-state Markov chain that 
the SR map is equivalent to. The transition probabilities are $p_{AA}=1/2, \, p_{AC}=1/2, \, p_{BA}=1, \, p_{CB}=1$, the rest being zero. The Markov matrix is then $T_1$ whose matrix elements are $t_{ij} p_{ij}$ while for fluctuations to be studied below the matrix $T_2$ whose elements are $t_{ij} p_{ij}^2$ is also useful. These matrices can be used to study the uniformity principle at the global as well as the local scales as is shown below.  To study what happens on restriction to smaller areas whose size can be controlled, as well as to find the actual locations of the orbits, it is useful to use a binary representation given ahead. This is not a symbolic dynamics in the sense that the dynamics is no more a left shift. However the dynamics  is an {\em eventual} left shift even in the binary representation and is used extensively below.

 Any point in phase space can be represented as a bi-infinite binary string $p.q$
representing its position $q$ and momentum $p$. The binary string representing the first $m$ bits of the position coordinate is labeled $\gamma_m$.  The quantity $m$ also corresponds to the number of times an orbit visits the stretching region (left half) of the square after some number of iterations.  The rules for the mapping equations on the binary string are given in~\cite{Lakshminarayan93} and summarized here: if the
most significant bit of position is a 0, the dynamics is that of a
left shift. If the most significant bit is a 1, position and
momentum coordinates are interchanged and in the \emph{new} momentum
coordinate all 0's and 1's are switched. For example, consider the period 3 orbit starting at
$(q_0,p_0) = (2/3,1/3)$: \beq (2/3,1/3) \rightarrow (2/3,2/3) \rightarrow (1/3,2/3) \rightarrow (2/3,1/3). \nonumber \eeq
The binary representations for the starting coordinates are $q = .\underline{10}$, $p = .\underline{01}$, where the underline indicates infinite repetition, and under the 
 dynamics this point maps as 
 \beq \underline{01}.\underline{10}  \rightarrow \underline{01}.\underline{01}  \rightarrow  \underline{10}.\underline{10}  \rightarrow 
\underline{ 01}.\underline{10} 
\eeq
 while under the symbolic dynamics this orbit is $\underline{CBA} \rightarrow \underline{BAC} \rightarrow \underline{ACB}.$
 
Making use of the symbolic dynamics, it is possible to prove that the map contains a dense set of
periodic orbits and is hence an ergodic transformation. It is also a straightforward argument to see that the positive Lyapunov exponent $\lambda_L$
for the SR map is \beq \lambda_L = \frac{1}{2} \ln 2. \label{lyapunov} \eeq This may be seen as follows:
The unique smooth invariant density is uniform  on the phase space. Therefore
 ergodicity implies that at a large number of iterations a typical orbit spends equal amounts of time on the left and right halves of the unit square. Since points along an orbit which are on the left half are stretched by a factor of two and points on the right half are not stretched at all, the Lyapunov exponent will be the average of $\ln 2$ and 0. The known symbolic dynamics, or binary representation, also allow the enumeration of all periodic orbits of any period as seen in Sect.~\ref{enumerate}.

\subsection{Periodic Orbits and Stability}
\label{sect:postability}

To begin the discussion of the periodic orbits, first note that there exist two exceptional periodic orbits on the boundary of the square: a period 1 fixed point at the origin and a period 2 orbit between the points $(1,1/2)$ and $(1/2,1)$.  All other orbits must pass through the interior of the rotating region $\mathcal R_4$ and it is thus sufficient to count orbits originating in $\mathcal R_4$. The orbits come in two types depending on the parameter henceforth called $j$, which is the number of changes from 0 to 1 or 1 to 0 in the binary string $\gamma_{m}$ representing the first $m$ bits of the position coordinate. If $j$ is odd, the orbit may be represented as $p.q = \underline{\gamma_{m}}. \underline{\gamma_{m}}$.  If $j$ is even the orbit may be written as $p.q = \underline{\gamma_{m}\overline{\gamma}_{m}}.\underline{\gamma_{m}\overline{\gamma}_{m}}$ where $\overline{\gamma}_{m}$ denotes the complement of $\gamma_{m}$. From the rules given for the symbolic dynamics one finds the period $\tau$ of an orbit in terms of its $\gamma_{m}$ string to be 
\beq 
\label{eqn:period} 
\tau = 2j + m + 2 
\eeq 
where $j$ may take any integer value from 0 to $m-1$. Reference~\cite{Lakshminarayan93} may be consulted for more details.

It is also useful to realize that it is possible to translate from the binary to the symbolic dynamics and vice-versa. Sticking to orbits originating from $\mathcal R_4$ these are of the form $\cdots 0.1x_2x_3\cdots$. This translates by replacing every transition (which is a 0-1 or 1-0 "bond" in the binary representation) including the initial 0.1 by $CBA$ and every other type ({\it i.e.} 0-0 and 1-1 bonds) by $AA$. Thus for example the orbit with the binary representation $\cdots 0.10100 \cdots$
translates to $\cdots CBACBACBACBAAA \cdots$.

The Jacobian stability matrix for a single time step is dependent
on whether the point in question is in the left or right half of the
unit square. Denoting ${\bf M}_{L}$ as the stability matrix for points in
the left half and ${\bf M}_{R}$ as the stability matrix for points in the
right half we have, following from the definition of the map, that \beq {\bf M}_{L} =
\begin{pmatrix}
             {2}  &  {0} \cr
             {0}  &  {1/2} \cr
   \end{pmatrix}, \hspace{2 mm} {\bf M}_{R} =
   \begin{pmatrix}
             {0}  &  {-1} \cr
             {1}  &  {0} \cr
   \end{pmatrix}. \label{leftrightjacobian} \eeq The  stability matrix along a trajectory Eq.~\refeq{stabilitymatrix}  may
   then be calculated explicitly as a product of the matrices Eq.~\refeq{leftrightjacobian}. Since
   points in $\mathcal R_4$ always map to $\mathcal R_3$ on the next time step, the product
   ${\bf M}_{R}{\bf M}_{R}$ will always come in pairs in the full product for
   ${\bf M}_{\tau}$. Because this product ${\bf M}_{R}{\bf M}_{R}$ produces a diagonal
   matrix, $-\bf{I}$, the full product Eq.~\refeq{stabilitymatrix} contains only diagonal matrices, so it is
    commutative. This implies that up to a minus sign $\bf M_{\tau}$ is determined by
   the number of times an orbit visits the left half of the unit
   square, and the parity is determined by whether $j$ is even
   or odd. Putting all of this together gives the period $\tau$ Jacobian stability
   matrix of a particular periodic orbit as \beq \label{eqn:jacobian} {\bf M}_{\tau} = \begin{pmatrix}
             {2^{m}(-1)^{j+1}}  &  {0} \cr
             {0}  &  {2^{-m}(-1)^{j+1}} \cr
   \end{pmatrix} \eeq 
with eigenvalues 
\begin{align}
 |\Lambda_{c}| &= 2^{-m}, \hspace{4 mm} \text{contracting} \nonumber \\
 |\Lambda_{e}| &= 2^{m}, \hspace{4 mm} \text{expanding}.
 \end{align}

 The explicit form of the inverse determinant that arises in semiclassical calculations is
 \beq
 \label{invdet}
\frac{1}{\left|{\rm Det}({\bf M}_\tau-{\bf 1})\right|} = \left\{ {\begin{array}{*{20}c}
   {(2^{m}  + 2^{ - m}  - 2)^{ - 1} {\text{,     j odd}}}  \\
   {(2^{m}  + 2^{ - m}  + 2)^{ - 1} ,{\text{     j even}}}  \\
 \end{array} } \right.
 \eeq
 which may be written as with a $\pm$ for notational convenience, knowing that the sign in front of the 2 is determined by the parity of $j$.  Since the main interest is in asymptotic calculations (large $\tau$) it is sufficient to keep the first two terms in the binomial expansion 
 \beq (2^{m} + 2^{ - m} \pm 2)^{ - 1} \approx 2^{-m} - ( - 1)^j 2^{ -2m + 1} + \cdots 
 \label{binapprox}
 \eeq 
It is shown in Sect.~\ref{sect:localdist} that actually only the first term in the expansion is necessary to investigate certain asymptotic fluctuation properties, leaving the approximation
\beq 
\frac{1}{\left|{\rm Det}({\bf M}_\tau-{\bf 1})\right|} \simeq 2^{-m}
\label{detapprox}
\eeq 
Thus, for large period, to leading order $\left|{\rm Det}({\bf M}_\tau-{\bf 1})\right|^{-1}$  coincides with the contracting stability eigenvalue $|\Lambda_{c}|$ of the stability matrix.  A final note, although the determinant is the quantity which arises in semiclassical theory, some of the classical dynamical systems literature defines the uniformity principle with respect to the inverse of the stretching exponential~\cite{Ott02}, in which case the correction term of Eq.~(\ref{binapprox}) is not relevant.

\subsection{Enumerating the periodic points}
\label{enumerate}

It is quite valuable to be able to count the periodic points of fixed period in a given subregion of the unit square.  To do so we proceed as follows: divide the unit square into a grid of boxes of area $2^{-k} \times 2^{-k}$ , whose lower left corners are specified by $q =.x_{1}x_{2}...x_{k}$ and $p = .y_{1}y_{2}...y_{k}$. This is a binary expansion, so each $x_{i}$ and $y_{i}$ is either 0 or 1. This is a total of $4^{k}$ boxes. Membership of an orbit in a box with a specified lower left corner is simply that the orbit has the same first $k$ bits for $p$ and $q$ in its binary representation as the lower left corner, and arbitrary bits beyond the $k$th. To start, consider boxes in the lower right subsquare $\mathcal R_4$.  The same  counting results will hold for regions $\mathcal R_3$ and $\mathcal R_2$ since these are merely rotations of $\mathcal R_4$. The slightly more detailed counting arguments for region $\mathcal R_1$ can be found in Appendix~\ref{appa}.

 As in Sect.~\ref{sect:postability} the periodic points are written
either in the form $\underline{\gamma_{m}}.\underline{\gamma_{m}}$
or
$\underline{\gamma_{m}\overline{\gamma}_{m}}.\underline{\gamma_{m}\overline{\gamma}_{m}}$,
the former if $j$, the number of 0-1 or 1-0 transitions in
$\gamma_{m}$, is odd, and the latter if $j$ is even. The requirement
on the string $\gamma_{m}$ for a point to be in $\mathcal R_4$ is that the first bit is 1.  If the last bit is 0, then the first form represents a point in $\mathcal R_4$, and if the last bit is 1, then it is the second form that is a point in $\mathcal R_4$. In either case, $m$ is related to the period $\tau$ by Eq.~\refeq{eqn:period}. Note that $j \leq m-1$, so $j \leq \tau - 2j - 3$ and $3j \leq \tau - 3$, so that $j$ is at most $\lfloor(\tau/3)\rfloor - 1$ where $\lfloor \cdots \rfloor$ denotes the floor function.  The first step is to count how many $\tau$-periodic points with a fixed value of $j$ there are in the box whose corner is specified by $q = .1x_{2}...x_{k}$ and $p =
.0y_{2}...y_{k}$ (note the 1 and the 0 are forced because the point must lie in $\mathcal R_4$).  This point is also represented by combining the binary expansions into one expression
$y_{k}y_{k-1}...y_{2}0.1x_{2}...x_{k}$, and similarly for other points.

For a $\tau$-periodic point of the first form, $\gamma_{m}$
must look like $1x_{2}...x_{k}w_{1}...w_{i}y_{k}...y_{2}0$ for some $i,k$ with the restriction $i
= \tau - 2j - 2k-2$ and the total number of transitions in this string
is $j$. If the periodic point is of the second form, then
$\gamma_{m} = 1x_2 ...x_k w_1 ...w_i \bar y_k ...\bar y_2 1$.

Let
$s$ equal the number of 0-1 or 1-0 transitions in $1x_2...x_k$ plus
the number of 0-1 or 1-0 transitions in $0y_2...y_k$. That is, $s$
is the total number of changes for the lower left corner point of
the box. If $s$ and $j$ are both odd, a periodic point would
be of the first form, and $\gamma_{m} =
1x_{2}...x_{k}w_{1}...w_{i}y_{k}...y_{2}0$. There are $s$
transitions from 1 to $x_k$ and from $y_k$ to 0 combined, so there
must be $j - s$ transitions in the $i+1$ possible places in
$x_{k}w_{1}...w_{i}y_{k}$. So there are
 $\left( {\begin{array}{*{20}c}
   {i + 1}  \\
   {j - s}  \\
 \end{array} } \right) = \left( {\begin{array}{*{20}c}
   {\tau - 2j - 2k - 1}  \\
   {j - s}  \\
 \end{array} } \right)$ ways to do this.
In the other cases in which $j$ and $s$ may be either even or odd,
the same result holds and thus we have that for each possible value of $j$, the number of
$\tau$-periodic points in a $2^{-k}$ $\times$ $2^{-k}$ box in $\mathcal R_4$ with corner
value specified by $s$ is $ \left( {\begin{array}{*{20}c}
   {\tau - 2j - 2k - 1}  \\
   {j - s}  \\
 \end{array} } \right) $.
The smallest possible value of $j$ is $s$, which occurs when all the
$w$'s are the same as $x_{k}$, and the maximum attainable value of
$j$ is $\lfloor(\tau - 2k + s -1)/3\rfloor$.  In addition, for a given value of $s$, there
are $\left( {\begin{array}{*{20}c}
   {2k - 2}  \\
   s  \\
 \end{array} } \right)$
 possible $2^{-k}$ $\times$ $2^{-k}$ boxes with $s$ the number of
transitions in the first $k$ $p$-bits plus transitions in the first
$k$ $q$-bits of the corner point.

To summarize, the counting just given is for $\tau$-periodic points in a binary grid of boxes within $\mathcal R_4$, $\mathcal R_3$, and $\mathcal R_2$ where $k$ is the number of bits specifying a box side, and $s$ is the number of transitions in the $k$ bits of the $q$-coordinate of the corner of the box plus the number of transitions in the $k$ bits of the $p$-coordinate of the corner point. The index $j$ ranges from $s$ to $\lfloor(\tau - 2k + s -1)/3\rfloor$ and for a given $j$ the number of periodic points with that value of $j$ is given by $\left( {\begin{array}{*{20}c}
   {\tau - 2j - 2k - 1}  \\
   {j - s}  \\
 \end{array} } \right) $, so the total number of period-$\tau$
points in this box is \beq N_{\tau}(s,k) = \sum_{j=s}^{\lfloor
(\tau-2k+s-1)/3\rfloor}{\VectorII{\tau-2j-2k-1}{j-s}}.
\label{eqn:numofpts1} \eeq Note that for fixed period and box size,
the statistics of the periodic points within a box are determined
entirely by the value $s$ of its corner point. Any two boxes with
the same value of $s$ will have exactly the same distribution, and
for each $s$ there are $\left( {\begin{array}{*{20}c}
   {2k - 2}  \\
   s  \\
 \end{array} } \right)$ such boxes. Thus, the Hannay-Ozorio  sum, Eq.~\refeq{HannayOzorio},  over all periodic points within a binary box in $\mathcal R_4$, $\mathcal R_3$, or $\mathcal R_2$ is 
 \begin{align}
 F_{\tau}(s,k) = \sum_{j=s}^{\lfloor (\tau-2k+s-1)/3\rfloor}&\VectorII{\tau-2j-2k-1}{j-s} \nonumber \\
 & \times \frac{1}{\left|{\rm Det}({\bf M}_\tau-{\bf 1})\right|} \label{hannaysum} 
 \end{align} 
 and the global form of Eq.~\refeq{HannayOzorio} (excluding $\mathcal R_1$) by summing over all boxes in $\mathcal R_4$, $\mathcal R_3$, and $\mathcal R_2$ gives 
 \begin{align} &F_{\tau} = 3\sum_{s=0}^{2k-2}\VectorII{2k-2}{s} F_{\tau}(s,k). 
 \label{eqn:R4sum} \end{align}
The relation of the inverse determinant to the period and the symbolic representation of an orbit is, from Eq.~\refeq{detapprox} and Eq.~\refeq{eqn:period}, given by 
\beq \frac{1}{\left|{\rm Det}({\bf M}_\tau-{\bf 1})\right|} \simeq 2^{-\tau+2j+2} 
\label{inversedet}
\eeq 
which provides an explicit summable expression for looking at fluctuations in the uniformity principle.

The counting arguments for the subsquare $\mathcal R_1$ are slightly different, but similar in character, to those presented here and the details are given in Appendix~\ref{appa}.  In fact, the resulting equations are quite close to the ones given in this section.

\begin{figure}[floatfix]
\includegraphics[width=0.45\textwidth]{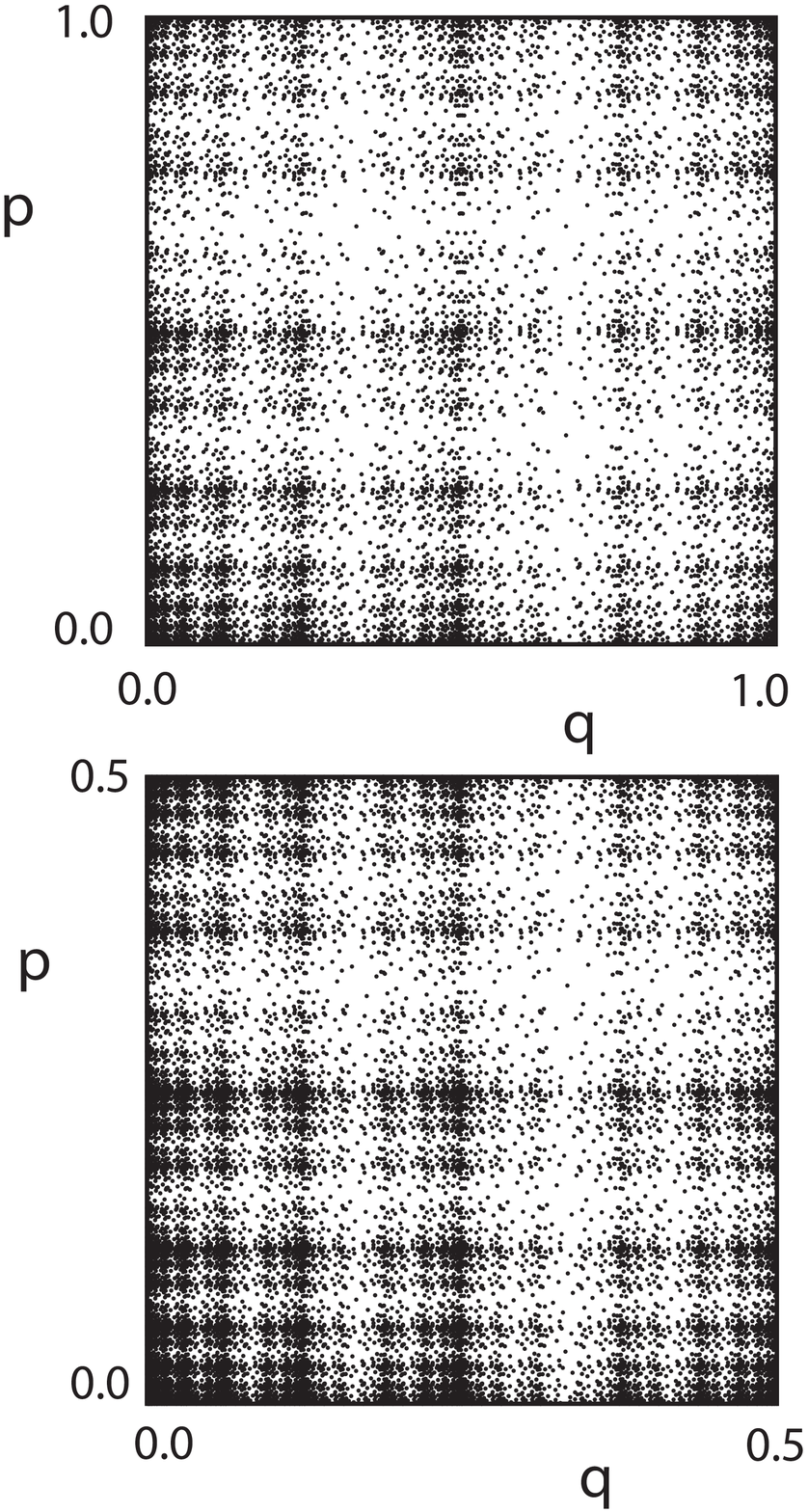}
\caption{Fixed points of the time $\tau$ iterated map.  The upper square is the plot for {$\tau$ = 24} and the lower square is the $R1$ region for {$\tau$ = 28}.   The density of fixed points at $\tau=28$ is roughly 4.6 times that of $\tau=24$.   Below, expanding $R1$ renders the lower plot's fixed point density similar to the upper plot.  The similarity of the expanded $R1$ fixed point's structure to the full phase space at an earlier time illustrates the fractal-like structure mentioned in the text.}
\label{fig:T28} 
\end{figure}

Figure~\ref{fig:T28} shows a plot of all periodic points in the unit square at $\tau=24$ and $28$. This visualization of the structure of the periodic points is interesting in its own right, as the points appear to have a fractal-like structure to them.  In fact the checkered pattern created mimics the stable and unstable manifolds of the map.

The symbolic dynamics and the Markov matrices can also be used to find the number of
periodic orbits as well as to study the uniformity principle. Since the SR map is simple enough to
permit both a combinatorial approach as well as a symbolic dynamics one, it is useful to 
present both. Given a periodic point of the first type in $\mathcal R _4$ whose $\gamma_m$ string has a binary representation $1x_2x_3\cdots w_1w_2\cdots w_i y_k \cdots y_2 0$, this translates into 
an orbit which is a repetition of  symbol strings of length $\tau= 2j+2k+i+2$. Of these 
$3(s+1)+[2(k-1)-s]=2s+2k+1$ are utilized to specify the fixed corner point. Thus there are 
$n=\tau-(2s+2k+1)$ number of possible ``free" symbols, say $S_1,\ldots S_n$. A little thought shows that the free symbol string {\it has} to end with $A$, that is $S_n=A$,  and {\it has} to be prefixed by an $A$. Thus it can be either of the form
$CBA \cdots A$ or $A \cdots A$. In both these cases the number of periodic points is then given by 
\beq
N_{\tau}(s,k)=\sum_{S_i} (T_0)_{AS_1} (T_0)_{S_1S_2} \cdots (T_0)_{S_{n-1}A},
\eeq
or
\beq
N_{\tau}(s,k)=\left(T_0^{\tau-2k-2s-1}\right)_{AA}.
\eeq
Recall that $T_0$ is the transition matrix in Eq.~\refeq{transition.matrix}. Thus the combinatorial problem
can be reduced to that of finding powers of a matrix. From this point the mathematical complexity is
comparable as both lead to the analysis of cubic equations; see Appendix~\ref{sect:sumformula} for the combinatorial case. 

The uniformity principle sum for the local area can also be written compactly in terms of a matrix 
power, this time the Markov matrix $T_1$. A similar reasoning as above leads to
\beq
F_{\tau }(s,k)=\dfrac{1}{2^{2k-1}}\left(T_1^{\tau-2k-2s-1}\right)_{AA}.
\label{symbdynFsk}
\eeq
Here however, the approximation in Eq.~\refeq{inversedet} is already used, as otherwise such
a compact formula is not possible. Note that with this approximation the global (over the whole phase space) uniformity principle sum is simply the trace of the power of the Markov matrix. That is
\beq
F_{\tau} = \mbox{Tr}(T_1^\tau).
\eeq
This follows from the fact that the entries in the stochastic matrix $T_1$ are precisely the 
multipliers which are either $1/2$ or $1$. That the stochastic matrix has necessarily an 
eigenvalue 1, and therefore $F_{\tau} \rightarrow 1$ as $\tau \rightarrow \infty$ is an alternative 
formulation of the uniformity principle. The other eigenvalues of $T_1$ whose eigenvalues are
less than 1 in modulus determine both the rate of decay of correlations as well as approach
to uniformity of the periodic orbits. We will expand on this below shortly. Almost all of the analysis 
below follows the consequences of the binary representation and the combinatorial approach
as  detailed statistics is more transparently done this way. 

\section{Statistical Results}
\label{sec:statistics}

The results of the previous section and Appendix~\ref{sect:sumformula} can be used to evaluate sum rule fluctuations.  Consider the local density of the inverse determinant $\left|{\rm Det}({\bf M}_\tau-{\bf 1})\right|^{-1}$, which occurs as the natural weighting for periodic orbits in many semiclassical expressions.  The first goal is to derive an asymptotic formula for its variance. This analysis leads naturally to discussing the density and convergence of the remaining component $\tilde F_{\tau}(s,k)$  of the Hannay-Ozorio sum rule introduced in Sect.~\ref{moments}.  We give an analytic expression for $\tilde F_{\tau}(s,k)$ and compute its variance, as well as local and global boundaries of convergence for the sum rule.

\subsection{Local distribution of the inverse determinant}
\label{sect:localdist}

Consider the regions $\mathcal R_4$, $\mathcal R_3$, and $\mathcal R_2$ (see \reffig{fig:regions}) of the unit square whose density of periodic orbits is described in Sect.~\ref{enumerate}.  As before,  the (similar) discussion for the region $\mathcal R_1$ is left to Appendix~\ref{appa}.  There is a range of values taken by $\left|{\rm Det}({\bf M}_\tau-{\bf 1})\right|^{-1}$ within a local patch of phase space, as described in Sect.~\ref{enumerate}.  It was shown that the number of period $\tau$ fixed points with a fixed value of the determinant specified by $j$ in a box with parameters $s$ and $k$ is given by  $\VectorII{\tau-2j-2k-1}{j-s}$ where $j \geq s$.  For large period the combinatorial as a function of allowed values of $j$ (which can be thought of as a probability density for various stability determinant values of fixed points in the box) is approximately normally distributed (see Appendix~\ref{sect:sumformula}) with an exact mean given by Eq.~\refeq{mean}, but written approximately here as  
\beq
\mu \simeq \frac{{\tau - 2s - 2k - 1}}{{5.148}}\,\,\, + \,\,\,s\,\, - 0.162. 
\eeq
Recall from Eq.~\refeq{inversedet} that at period $\tau$ the inverse determinant may take on the values $2^{-\tau + 2j +2}$ as $j$ ranges from $s$ to $\lfloor(\tau - 2k + s - 1)/3\rfloor$.  So in fact, for the discussion of $\left|{\rm Det}({\bf M}_\tau-{\bf 1})\right|^{-1}$, the sum rule $F_\tau(s,k)$ is over terms of the form $\VectorII{\tau-2j-2k-1}{j-s}4^j$.  It turns out that the density for these terms is normal as before for the $t=0$ case; note, oddly enough,  that does not imply that the density for finding a particular value of the  inverse determinant is lognormal as the convergence with $\tau-2j-2k-1\rightarrow\infty$ to normal is too slow.

Consider the density $g(i,t) = \VectorII{n-2i}{i}e^{ti}$ as a function of $i$ for a given $t$, where as in Appendix~\ref{sect:sumformula}, $e^t = \alpha$.  Using Stirling's formula to approximate $g(i,t)$, and calculus, one finds that the maximum value of $g(i,t)$ occurs at the value $i_0$ of $i$ given by
\beq
 i_0 \approx n \frac{\beta_1(t) - 1}{3 \beta_1(t) - 2} 
 \eeq 
 where $\beta_1(t)$ is the real root of the cubic equation $ \beta^3 - \beta^2 - e^t = 0$. Interestingly this same cubic equation arises here for a different problem from the one considered in Appendix~\ref{sect:sumformula}.  This method does not give the exact  transient terms as the recurrence method of Appendix~\ref{sect:sumformula} does, but the same structure exists and near the maximum at $i_0$, $g(i,t)$ is approximated continuously as a Gaussian with width on the order of $\sqrt{n}$ and so the values of $i$ which contribute to the sum are sharply peaked around the maximum. Specifically,
\begin{align}
g(i,t) &= \left( {\begin{array}{*{20}c}
     {n - 2i}  \\
     i  \\
  \end{array}} \right)e^{ti}  \nonumber \\
  &\simeq
  \frac{\beta_1^{n+1} }{{3\beta_1  - 2}} \frac{1}{{\sqrt {2\pi \sigma^2_i} }}e^{ - \,(i - i_0 )^2 /(2\sigma^2_i)}
  \end{align} 
  where $\sigma^2_i = n \frac{{\beta_1 (\beta_1  - 1)}}{{(3\beta_1  - 2)^3 }}$.

For the quantity $\left|{\rm Det}({\bf M}_\tau-{\bf 1})\right|^{-1}$, for which $e^t = 4$, $\beta_1 = 2$ and so the  mean of $g(i,t)$ occurs at $i_0 = n/4$, even though the mean $\mu(n)$ (for the unweighted combinatorial) occurs at about $n/5.148$.  As $n \rightarrow \infty$, the two densities tend toward a vanishing overlap since the difference in the means grows faster than the widths.  In considering values taken by the inverse determinant, only those periodic points with transition number $j$ which occur near 
\beq 
j_0 = \frac{\tau-2s-2k-1}{4} + s \label{j0det} 
\eeq
contribute to the sum $F_{\tau}(s,k)$, in spite of the fact that there is a vanishing relative fraction of fixed points associated with this value of $j$, as most points have a value of $j$ near $(\tau-2s-2k-1)/5.148 + s - 0.162$.

\subsection{Important Moments}

We give here explicit expressions for some of the quantities of interest related to the inverse determinant and the SR map, using the results from the methods of Appendix~\ref{sect:sumformula}.  First consider the number of period $\tau$ fixed points within a binary box specified by $s$ and $k$.  The number $N_{\tau}(s,k)$ is given by Eq.~\refeq{eqn:numofpts1},  which is a special case of the sum formula $S(n,\alpha)$ from Appendix~\ref{sect:sumformula} with $\alpha = 1$ and $n = \tau - 2k - 2s - 1 $. The form of the solution is therefore specified by Eq.~\refeq{genericform}, with appropriate values for the constants. The topological entropy is given by $h_T=\ln \beta_1(0)$, and thus  
\beq 
N_{\tau}(s,k) = c_1 e^{h_T n} \, + 2e^{-h_T n/2}\left[ a\cos(n\theta) - b\sin(n\theta) \right]
\eeq
This equation contains the finite-time correction terms to the count of fixed points of a binary box, which cannot be given by specifying the entropy alone.  If $\tau$ is large, the leading term $c_{1}e^{h_T n}$ dominates and may be used for asymptotic calculations.

The moments for the inverse determinant (which are different from the moments involved in the sum rule fluctuations, ahead) may be computed by averaging over powers of $\left|{\rm Det}({\bf M}_\tau-{\bf 1})\right|^{-1}$. The most important case is the mean, which corresponds to $F_{\tau}(s,k)$, and gives an explicit finite-time correction term to the (infinite time) prediction of the uniformity principle.  It is also an ingredient of other sum rule moments.

From Eq.~\refeq{hannaysum} and Eq.~\refeq{inversedet}, the local form of the sum of the inverse determinant over all fixed  points within a box reads 
\beq 
F_{\tau}(s,k) = \sum\limits_{j = s}^{\left\lfloor {(\tau - 2k + s - 1)/3} \right\rfloor } {\left(
{\begin{array}{*{20}c}
   {\tau - 2j - 2k - 1}  \\
   {j - s}  \\
\end{array} } \right)} 2^{ - \tau + 2j + 2}
\label{eqn:Fs} 
\eeq
 which is predicted by the Hannay-Ozorio  sum rule to asymptotically approach the area of the box, $4^{-k}$.  A slight change of variables puts the sum in the generic form, Eq.~\refeq{appendixform}, with $\alpha = 4$ and $n = \tau - 2k - 2s -1$.  The solution is thus again of the form of Eq.~\refeq{genericform}, where the real root $\beta_1$ of the cubic is exactly 2.  After some algebraic manipulation, it is seen that $F_{\tau}(s,k)$ may be written in a form which displays both its exponential dependence on the period as well as its relation to the phase space area  as
\beq 
F_{\tau}(s,k) = 4^{-k} + e^{ - \lambda_L \tau} 2^{s - k + 5/2} [a \cos (n\theta ) - b \sin (n\theta)]
\label{eqn:Fs2} 
\eeq
where $\lambda_L $ is the real part of the leading Pollicot-Ruelle resonance (here also equal to the positive Lyapunov exponent) whose value is given in Eq.~\refeq{lyapunov}. The numerical values of the constants $a$, $b$, and $\theta$ may be calulated using the formulas of Appendix.~\ref{sect:sumformula}.  Subtracting the Hannay-Ozorio term leaves the oscillating part with time of the sum rule as 
 \beq 
 \tilde F_{\tau}(s,k) = e^{ - \lambda_L \tau} 2^{s - k + 5/2} [a \cos (n\theta ) - b \sin (n\theta)]. \label{eqn:fluctuating}
 \eeq
By factoring out the time dependence, it turns out that the rate of convergence towards uniformity with increasing period is exponential, as expected.  It is clear from the discussion  based on symbolic dynamics and Eq.~\refeq{symbdynFsk} that the rate is governed by the eigenvalues of a Markov matrix if it exists or generally by the Pollicott-Ruelle resonances.  In the case of the SR map,
the modulus of the eigenvalues of the Markov matrix, which are the resonances, have modulus
equal to $1/\sqrt{2}$, which is $e^{-\lambda_L}$.  It is well-known that the Pollicott-Ruelle resonances are generically not related to the Lyapunov exponents and therefore the equality for the SR map must be considered a coincidence. The close connections between mixing and uniformity principle makes the emergence of the Pollicott-Ruelle resonances as governing the rate of convergence to uniformity much more natural. 

It is also interesting to consider the convergence boundary as mentioned in Sect.~\ref{sect:averages}.   This amounts to determining the box size (phase space volume) for a given period and location in phase space at which the size of the correction term $\tilde F_{\tau}(s,k)$ is just the same order of magnitude as the local area itself.  In particular, from Eq.~\refeq{eqn:fluctuating}, if $2^{-\tau/2 + s - k + 5/2} = 4^{-k}$ then $k = \tau/2 -s -5/2$ and the volume at the convergence boundary is 
\begin{equation} 
\mathcal{V} (s;\tau) = 2^{-\tau + 2s + 5}. 
\end{equation} 
In this way,  the local sum rule fluctuations are equally as important as the mean, and hence to any results which invoke a sum rule on that local scale at that time.  Given that $s$ varies in the domain $0\le s \le 2k-2$ or in terms of $\tau$, $0 \le 3s \le 2\tau -7$, the convergence boundary varies greatly from one location to another in the phase space, i.e. $2^{-\tau} \le \mathcal{V} (s;\tau) \le 2^{-\tau/3}$.  Although the local convergence boundary vanishes everywhere as $\tau\rightarrow \infty$, its relative variation tends to infinity.  The relatively larger boundaries are precisely linked to locally greater inverse determinant variation just ahead.

In Eq.~\refeq{binapprox}, the leading correction term to  the inverse determinant was given, but up to this point not included in the calculations.   It is important to know if this error is subdominant relative to the fluctuating component just calculated.  If the second expansion term is kept, this leads to a sum denoted $\tilde B_{\tau}(s,k)$ of the form 
\begin{align} 
\tilde B_{\tau}(s,k)  = \sum\limits_{j =s}^{\left\lfloor {(\tau - 2k + s - 1)/3} \right\rfloor } &{\left(
{\begin{array}{*{20}c}
   {\tau - 2j - 2k - 1}  \\
   {j - s}  \\
 \end{array} } \right)} \nonumber \\ 
 & \times 2^{ - 2\tau + 5}(-16)^{j} 
 \label{sumcorrection}
  \end{align} 
  which once again is in the form of the sum discussed in Appendix~\ref{sect:sumformula} with $\alpha =
 - 16$ and $n = \tau -2s - 2k -1$.

In fact, more properly, the two most dominant corrections to the local sum rule are
\beq 
F_{\tau}(s,k) = 4^{-k} + \tilde F_{\tau}(s,k) - \tilde B_{\tau}(s,k) 
\eeq 
We know that the first correction term is governed by the Pollicott-Ruelle resonances, but the second term is something else.  A priori, it is not obvious which of these two correction terms dominates for large period.  Extracting only the exponential dependence on $\tau$ gives $\tilde F_{\tau}(s,k) \propto e^{-\lambda_L \tau} \approx (0.71)^\tau$.  For $\tilde B_{\tau}(s,k)$, because $\alpha<0$, the dominant fluctuation term comes from the oscillatory $(\alpha/\beta)^{n/2}$, which is approximately $\tilde B_{\tau}(s,k) \approx (0.67)^\tau$.  In this case, the first correction term eventually dominates over the second, and using the approximation of Eq.~\refeq{inversedet} is justified.  Had the situation turned out the opposite way, then the correction term would not have been a Pollicott-Ruelle resonance; we are not aware of an argument suggesting that this could not have happened and thus both sources of corrections must be considered in other cases.

Before continuing with the spatial fluctuations in the sum rule itself, consider the variation of the individual inverse determinants contributing to each sum.  They vary wildly from one fixed point to the next and there is greater variation in some regions as opposed to others.  This gives an $s$-dependence to their variation within any single box.  This can be seen by computing the variance.  The sum of squares of the inverse determinant, $Q_{\tau}(s,k)$, is given by
 \begin{align} 
 Q_{\tau}(s,k) &= \sum_{f.p.} \frac{1}{\left|{\rm Det}({\bf M}_\tau-{\bf 1})\right|^2} \nonumber \\
 &= \sum\limits_{j =
s}^{\left\lfloor {(\tau - 2k + s - 1)/3} \right\rfloor } {\left(
{\begin{array}{*{20}c}
   {\tau - 2j - 2k - 1}  \\
   {j - s}  \\
 \end{array} } 
 \right)} 2^{ - 2\tau + 4j + 4} 
 \end{align} 
This is a sum of the form of Eq.~\refeq{appendixform} with $\alpha = 16$ and $n = \tau - 2k - 2s -1$. Keeping only the leading term in Eq.~\refeq{genericform},  the solution is 
\beq Q_{\tau}(s,k) =
 4^{-\tau+2s+2}c_{1Q}\beta_{1Q}^{\tau-2k-2s-1} 
 \eeq 
 where $\beta_{1Q}$, the real root of the cubic polynomial for $\alpha = 16$, is approximately 2.901 and $c_{1Q} \approx 0.433$. The subscripts on the numerical constant $c$ and $\beta$ are used here to make it clear that these refer to the case $Q$, for which $\alpha = 16$. The same result can also be derived from the symbolic dynamics as 
 \beq
 Q_{\tau}(s,k)=\dfrac{1}{4^{2k-1}} \left(T_2^{\tau-2k-2s-1}\right)_{AA}.
 \eeq

In the notation of Eq.~\refeq{moment}, the variance of the inverse determinants within a given box is $\sigma^2(e^{\lambda\tau}) = {\cal M}_{-2}(  e^{\lambda\tau},s,k) - {\cal M}_{-1}(  e^{\lambda\tau},s,k)^2$ and
\beq 
\sigma^2(e^{\lambda\tau},s,k) = \frac{Q_{\tau}}{N_{\tau}} - \left( \frac{4^{-k}}{N_{\tau}}\right)^{2} \approx \frac{Q_{\tau}N_{\tau} - 2^{-4k}}{N_{\tau}^{2}} 
\eeq 
where the leading order of the mean is sufficient.  The product $Q_{\tau}N_{\tau}$ depends on $\tau$ by a factor $(e^{h_T}\beta_{1Q}/4)^{\tau}$, which is greater than unity.   Thus, this term diverges as $\tau\rightarrow\infty$. Asymptotically the local variance is just $Q_{\tau}/N_{\tau}$, or 
\begin{equation}
\sigma^2(e^{\lambda\tau},s,k) \longrightarrow 
  \frac{16c_{1Q}}{c_{1N}}\left( {\frac{{\beta_{1Q} }}
{{4e^{h_T}}}} \right)^{\tau-2s} \left(
{\frac{{\beta_{1Q} }} {{e^{h_T}}}} \right)^{ - 2k - 1} 
 \end{equation} 
 which shows asymptotically how the variance varies with $s$, a local characteristic of a particular region of phase space (box). Here $\beta_{1Q} / e^{h_T}$ is about 1.98 and $\beta_{1Q} / 4 e^{h_T}$ is about 0.495. When the analogous details are worked out for boxes in the region $\mathcal R_1$, the variance differs only by a constant factor of $16 \left(\frac{\beta_{1Q}}{e^{h_T}}\right)^{2}$.

Note that boxes whose lower-left corner has a small number of transitions (small $s$) have smaller variances, as well as more $\tau$-periodic points, than boxes with large $s$.   Precisely as found for the local convergence boundaries,  the variation of $s$ for moderately large $k$ leads to an enormous difference in the variations within different boxes of the same size.  Although, the variances vanish in the limit of $\tau\rightarrow\infty$, the ratios of the variances from one box to another another increase indefinitely as the box size shrinks.  

 \subsection{Sum rule fluctuations}
 \label{sumfluct}
 
 Next the global variance of the local sum rule $F_{\tau}(s,k)$ is considered due to spatial variation.  First we comment on the form of the density of $F_{\tau}(s,k)$.  Recall that with the method of subdividing the phase space into a grid of binary boxes, the value of $F_{\tau}(s,k)$ locally within a box is specified by a parameter $s$ which counts the number of 0 to 1 changes in the binary representation of the lower left corner of the box (see Sect.~\ref{enumerate}).  Thus, $\tilde F_{\tau}(s,k)$ depends exponentially on $s$, Eq.~\refeq{eqn:fluctuating}.   Furthermore, the number of boxes throughout a quarter region of the unit phase space square with a given value of $s$ is $\VectorII{2k - 2}{s}$, as $s$ ranges from 0 to $2k-2$.  Thus, the density of the logarithm of $\tilde F_{\tau}(s,k)$ follows a binomial centered at $k-1$.  As with the inverse determinant, the distribution of  $\tilde F_{\tau}(s,k)$ is described by the product of an exponential function (of $s$, here) and a combinatorial coefficient that is approximately normal.  A qualitatively similar behavior to the discussion of Sect.~\ref{sect:localdist}  arises in describing the density of values taken by the sum formula $\tilde F_{\tau}(s,k)$.

The variance of $\tilde F_{\tau}(s,k)$ is the average square deviation from the mean summing over all boxes and dividing by their total number, $4^{k}$.  This is essentially the second moment defined in Sect.~\ref{moments}
\beq 
{\cal M}_2(\tilde F_{\tau},k) = \frac{1}{4^k} \sum_{s=0}^{2k-2}
\left( \begin{array}{c}
   2k - 2 \\
   s  \\
\end{array}  \right) 
\tilde F_{\tau}(s,k)^2
\label{variance2} 
 \eeq
For calculational convenience, Eq.~\refeq{eqn:fluctuating} is rewritten in the form 
\beq 
\tilde F_{\tau}(s,k) =  Ae^{\gamma s} + (Ae^{\gamma s})^*
\label{Fgamma} 
\eeq 
where $\gamma = \ln{2} - 2\theta i$ and $A = c_{2}2^{-\tau /2 - k +3/2}e^{i\theta (\tau - 2k - 1)}$. Recall that the constants $c_2$ and $\theta$ arise from the solutions of the sum formula in Appendix~\ref{sect:sumformula}, in this case for $\alpha = 4$.  This result holds for the regions $\mathcal R_4$, $\mathcal R_3$, and $\mathcal R_2$ of the unit square. For $\mathcal R_1$, the expression used for $F_{\tau}(s,k)$ differ only by a constant factor, as  shown in Appendix~\ref{appa}, and this factor is accounted for below in giving the variance over the entire unit square.

For large $k$, it is possible to find a simplified asymptotic expression for the variance.  Let $c_2=|c_2|e^{i\zeta}$ and $\eta = \theta(\tau - 2k - 1) + \zeta$ giving
\beq 
[Ae^{\gamma s} + (Ae^{\gamma s})^*]^2 = |c_2|^2 2^{-\tau - 2k + 3}[2^{2s+1} + 2 \text{Re}(e^{2i\eta}e^{2\gamma s})].
\eeq 
The expression for the variance becomes 
\begin{align} 
{\cal M}_2(\tilde F_{\tau},k) =  & \hspace{1 mm} 4^{-k}|c_2|^2 2^{-\tau - 2k +4} \left[\sum_{s=0}^{2k-2}{\left( {\begin{array}{*{20}c}
   {2k - 2}  \\
   s  \\
 \end{array} } \right)4^s}\right.  \nonumber \\
  & \hspace{5 mm} \left. + \text{Re}(e^{2i\eta}\sum_{s=0}^{2k-2}{\left( {\begin{array}{*{20}c}
   {2k - 2}  \\
   s  \\
 \end{array} } \right)e^{2\gamma s}})\right]. 
 \end{align} 
Recalling the binomial theorem, each term above may be summed explicitly to give 
\begin{align} 
{\cal M}_2(\tilde F_{\tau},k) = &|c_2|^2 2^{-\tau}16^{-k+1} \left[ 5^{2k-2} \right.
\nonumber \\
  & \hspace{5 mm} \left. + \text{Re}\left(e^{2i\eta}\left[1+e^{2\gamma}\right]^{2k-2}\right)\right]. 
  \end{align} 
Letting $1+e^{2\gamma} = 1 + 4e^{-i4\theta} = \rho e^{i\omega}$ where $\rho^2 = 17 + 8\cos(4\theta)$  gives
\begin{align} 
{\cal M}_2(\tilde F_{\tau},k) &= |c_2|^2 2^{-\tau-4k+4}\left[5^{2k-2} + \rho^2 \cos(2\eta + 2\omega[k-1])\right] \nonumber \\ 
&= |c_2|^2 2^{-\tau}\left(\frac{25}{16}\right)^{k-1} \nonumber \\
   & \hspace{4 mm} \times\left[1 + \left(\frac{\rho^2}{25}\right)^{k-1} \cos(2\eta + 2\omega[k-1])\right]. 
   \end{align} 
Since $4\theta$ is not a multiple of $2\pi$, $\cos(4\theta)$ is less than unity and so is $\rho^2 / 25$.  Thus, the oscillatory terms are subdominant as $k$ increases.  For large $k$, or small local volume, the expression for the variance in each of regions $\mathcal R_4$, $\mathcal R_3$, and $\mathcal R_2$ becomes:  
\beq 
{\cal M}_2(\tilde F_{\tau},k) \simeq (3/4)|c_2|^2 e^{-2\lambda_L \tau}(5/4)^{2k-2}. 
\eeq 
For the region $\mathcal R_1$ the same equation for $\tilde F_{\tau}(s,k)$ as Eq.~\refeq{Fgamma} applies except that the coefficient $A$ is replaced by $A' = c_{2}2^{-\tau /2 - k +1/2}e^{i\theta (\tau - 2k + 1)}$. From this it follows that the contribution to the variance from the region $\mathcal R_1$ is simply one fourth the value for $\mathcal R_4$, $\mathcal R_3$, or $\mathcal R_2$.  The asymptotic formula for the variance of $F_{\tau}$ taken over the entire unit square is 
\beq 
{\cal M}_2(\tilde F_{\tau},k) \simeq \frac{13}{16} |c_2|^2 e^{-2\lambda_L \tau}\left(\frac{5}{4}\right)^{2k-2}. 
\label{eqn:variance} 
\eeq 
The variance thus decreases exponentially with time, again governed by the Pollicott-Ruelle resonance.  It also increases with the decreasing local volumes.  This gives a global convergence boundary for the sum rule on which the variance over the entire phase space remains a constant (rather than vanishing).  From Eq.~\refeq{eqn:variance} this would be given approximately by $k = \frac{\lambda_L\tau}{\ln{5/4}}$, and 
\beq 
\mathcal{V} (\tau)  = 2^{-\tau\ln{2}/\ln{5/4}} \approx 2^{-3.1 \tau}. 
\eeq

\section{Concluding Remarks}

The convergences and fluctuations of classical sum rules are interesting in a multitude of ways.  Although, their corrections may be exponentially suppressed with increasing time, the individual contributions can have a diverging variance themselves.  Another interesting feature of local sum rules, as  shown herein, is that certain fluctuations can be surprisingly large as the location of phase space is varied.  Correction terms may be related to known properties of the system in more general dynamical systems, such as  the topological entropy, the Pollicott-Ruelle resonances, or the Lyapunov exponent depending on the precise sum rule of interest.  It would appear that a fluctuation quantity, which depends sensitively on some higher power of the stability determinant, if such a quantity exists, may be more likely to reflect the kinds of fluctuations that have been described here on a theoretical basis for the SR map.  The results, however, may be suggestive of the type of behavior one might expect in a regime where sum rule fluctuations could arise. The asymptotic form of several different fluctuation measures derived for the Hannay-Ozorio sum, as well as their time and length scales, came from the solution of the same simple cubic polynomial. The origin of this lies in the symbolic dynamics which is a subshift of finite type on three symbols and thus there is an equivalent three state Markov chain. It may be the case that similar methods could be applied to other relatively simple systems, and that sum rule corrections could also be derived for these systems from the basis of their dynamics. 

The main results for the SR map begin with the calculation of the two sources of fluctuations in the Hannay-Ozorio sum rule.  The first source is governed by the dominant Pollicott-Ruelle resonances.  It arises from the non-uniformity of the locations of fixed points and their non-uniform weighting by the leading behavior of their inverse stability determinants.  The second source arises from the effects of next-to-leading order corrections to the inverse stability determinants.  These corrections are not governed by the Pollicott-Ruelle resonances, but are also exponentially decreasing in time.  The dominant correction here comes from the first source and hence the Pollicott-Ruelle resonances, however we do not currently know whether this must be the case for general chaotic dynamical systems. 

It is a matter of how closely the stretching multipliers approximate the determinant in Eq.~(\ref{invdet}), and it could be that for some other chaotic system they are different enough to produce corrections that dominate the one due to the resonances, although these would still be present. In the specific case of the SR map  the second term in Eq.~(\ref{binapprox}), the principal correction, is an oscillating  sum  because the periodic orbits are reflecting hyperbolic if the number of $0-1$,$1-0$ bonds are odd. This term can be written as a trace of the power of the matrix
\begin{equation}
\left( \begin{array}{ccc}1/4&0&1/4\\-1&0&0\\0&1&0 \end{array} \right)
\end{equation}
which is different from the $T_2$ matrix in that the element $(2,1)$ is $-1$ rather than $1$. This
ensures that each time the orbit gets rotated, it acquires a negative sign. Alternately, for each CBA part of the symbolic
string a negative sign is acquired. 

The leading eigenvalue of this matrix has a modulus of $\approx 0.67$ 
which is smaller than the subleading eigenvalue, $\approx 0.707$, of
$T_1$ that gives the Pollicot-Ruelle resonance. If the orbit were {\it all} direct hyperbolic then in the above matrix the $-1$ will change to $1$ and this will be same as $T_2$ whose leading eigenvalue is
$0.725$ which is larger than $0.707$, and would have dominated the corrections.
The fact that some of the orbits are reflecting hyperbolic seems to have been crucial to lower the contribution from corrections that come from the fact that a $\det(J-I)$ is present instead of just the
multipliers.

However if the sum rule is weighted by (the inverse) of the largest eigenvalue of the stability matrix, the Pollicott-Ruelle resonances will 
govern the corrections to the sum rule, especially if there is a finite symbolic dynamics 
description of the system. The relevance of Markov and related matrices ($T_0,T_1,T_2, \ldots$)            
 for the calculation of the fluctuations indicate possible connections with the Thermodynamic Formalism especially as applied to finite Markov processes \cite{Gaspard98}. 

 A second result shows how the relative local variations of the inverse determinants varies infinitely broadly at long times.  Finally, the relative variation of the sum rule applied locally also has an infinite width while maintaining an exponential convergence rate for fixed phase space volume.  We gave convergence boundaries that show how small a local volume may be considered for a given time of propagation if one expects convergence to the asymptotic sum rule result.  Again, the relative size of a converged local volume depended on location and varied infinitely broadly while maintaining exponential convergence with time at fixed volume
  
It would be extremely interesting to investigate other sum rules, especially those that connect to quantum fluctuation properties of eigenfunctions and transport.  The various localizing effects giving rise to eigenfunction scarring~\cite{Heller84}, localization manifestations of time scales introduced by transport barriers~\cite{Bohigas93}, and interaction effects linked to Friedel oscillations~\cite{Tomsovic08, Ullmo09} give a few interesting directions for further studies.  As mentioned earlier, the SR map is easily studied quantum mechanically and would be one possible way to study sum rules arising from quantum fluctuations properties involving eigenfunctions.

\begin{acknowledgments}
We would like to acknowledge very helpful discussions with J.~H.~Elton on several of the mathematical points involved, and the generous support of the U.S.~National Science Foundation grants PHY-0855337 and PHY-0649023.
\end{acknowledgments}

\appendix

\section{Region $\mathcal R_1$} 
\label{appa}

 Here the counting arguments and several results for the region $\mathcal R_1$ of the unit square, which has mostly been ignored in the body of the text, are presented. The reason for leaving this discussion here is that many of the derived results closely resemble those for the other regions, although the arguments are somewhat longer.

We begin with an extension of Sect.~\ref{enumerate} by counting the number of period $\tau$ points in a binary $2^{-k}$ by $2^{-k}$ box in the region $\mathcal R_1$ (\reffig{fig:regions}), but not on the bottom row of boxes. The lower left corner of each box is defined by $\underset{\raise0.3em\hbox{$\smash{\scriptscriptstyle-}$}}{0} y_k...y_2 0.0x_2 ...x_k
\underset{\raise0.3em\hbox{$\smash{\scriptscriptstyle-}$}}{0} $, with the condition that not all of the $y$'s are zero. The upper right corner is given by
$\underset{\raise0.3em\hbox{$\smash{\scriptscriptstyle-}$}}{1} y_k...y_2 0.0x_2 ...x_k
\underset{\raise0.3em\hbox{$\smash{\scriptscriptstyle-}$}}{1} $. It is easy to see that after applying the inverse transformation of the map some number of times, each square of area $4^{-k}$ will be mapped
into a rectangle in $\mathcal R_4$ of the same area, although not square, and also that the upper right corner of the square in $\mathcal R_1$ gets mapped into the lower left corner of the rectangle in $\mathcal R_4$. Thus the box in $\mathcal R_1$  with lower-left corner $y_{k}...10...0 . 0x_2...x_k$ (where not all the $y$'s are zero)  has upper right corner
$\underset{\raise0.3em\hbox{$\smash{\scriptscriptstyle-}$}}{1} y_k
...10...0\,.\,0x_2 ...x_k
\underset{\raise0.3em\hbox{$\smash{\scriptscriptstyle-}$}}{1} $
which transforms under the inverse transformation as follows:
$\underset{\raise0.3em\hbox{$\smash{\scriptscriptstyle-}$}}{1} y_k
...10...0\,.\,0x_2 ...x_k
\underset{\raise0.3em\hbox{$\smash{\scriptscriptstyle-}$}}{1}
\leftarrow
\underset{\raise0.3em\hbox{$\smash{\scriptscriptstyle-}$}}{1} y_k
...1.0...\,0x_2 ...x_k
\underset{\raise0.3em\hbox{$\smash{\scriptscriptstyle-}$}}{1}
\leftarrow
\underset{\raise0.3em\hbox{$\smash{\scriptscriptstyle-}$}}{0} \bar
x_k ...\bar x_2 1...1\,.\,1...y_k
\underset{\raise0.3em\hbox{$\smash{\scriptscriptstyle-}$}}{1}
\leftarrow
\underset{\raise0.3em\hbox{$\smash{\scriptscriptstyle-}$}}{0} \bar
y_k ...0\,.\,1...1\bar x_2 ...\bar x_k
\underset{\raise0.3em\hbox{$\smash{\scriptscriptstyle-}$}}{0} $, so
the lower left corner of the rectangle in $\mathcal R_4$ is $\bar y_k
...0\,.\,1...1\bar x_2 ...\bar x_k $, with transition number one less than the lower left corner of the box in $\mathcal R_1$.

Let $s$ be the number of transitions of the lower left corner of a box in $\mathcal R_1$.  Then by the same counting argument that was used before (the fact that it is a rectangle instead of a square does not change things), the number of $\tau$-periodic points in this box in $\mathcal R_1$ with a given
transition number $j$ is $\left( {\begin{array}{*{20}c}
   {\tau - 2j - 2k - 1}  \\
   {j - s + 1}  \\
 \end{array} } \right)$
 as $j$ ranges from $s-1$ to
$\lfloor(\tau - 2k + s -2)/3\rfloor$. So this looks just like it did for the $\mathcal R_4$ case except $s$ is replaced by $s$ - 1.

Now, the lower left corner of a box that we are considering in this region has, say, $t$ transitions in the position coordinate and $v$ transitions in the momentum coordinate, where $0 \leq  t \leq k-1$, and $1 \leq v \leq k-1$, and $s = t + v$.  The combinatorial expression above shows that the number of $\tau$-periodic points in a box depends only on the total $s$ and not how it is distributed between $t$ and $v$, however it is necessary to consider $t$ and $v$ separately because it is $v$ that is restricted to be greater than zero and not just the sum of $t$ and $v$.  We separately choose $t$ transitions from $k-1$ places for position, and $v$ transitions from $k-1$ places for momentum, with $v$ restricted to be greater than zero.  So, for example, the local form of the sum of the inverse determinant for boxes in $\mathcal R_1$ excluding the bottom row (denoted $\mathcal R_{1a}$), analogous to expression Eq.~\refeq{hannaysum}, is 
\begin{align} F_{\tau}(t+v,k;\mathcal R_{1a}) =
 \sum\limits_{j = t + v - 1}^{\left\lfloor {(\tau - 2k + t + v - 2)/3} \right\rfloor } &{\left( {\begin{array}{*{20}c}
   {\tau - 2j - 2k - 1}  \\
   {j - (t + v - 1)}  \\
 \end{array} } \right)} \nonumber \\
  & \times \frac{1}{\left|{\rm Det}({\bf M}_\tau-{\bf 1})\right|} 
  \label{R1local}
  \end{align}
and the sum over all boxes in $\mathcal R_{1a}$, analogous to expression Eq.~\refeq{eqn:R4sum}, is
 \beq _1F_{\tau a} =
\sum\limits_{v = 1}^{k - 1} \sum\limits_{t = 0}^{k - 1} \left(
{\begin{array}{*{20}c}
   {k - 1}  \\
   v  \\
 \end{array} } \right)\left( {\begin{array}{*{20}c}
   {k - 1}  \\
   t  \\
 \end{array} } \right) F_{\tau}(t+v,k;\mathcal R_{1a}).
 \eeq
Considering the bottom row of boxes in $\mathcal R_1$ (denoted $\mathcal R_{1b}$) a similar, but more tedious argument, gives the number of period $\tau$ points again as $\left( {\begin{array}{*{20}c}
   {\tau - 2j - 2k - 1}  \\
   {j - s + 1}  \\
 \end{array} } \right)$ as this time $j$ ranges from $s$ to $\lfloor(\tau-2k+s-2)/3\rfloor$. So the local sum of the inverse determinant for boxes on the bottom row of $\mathcal R_1$ is 
 \begin{align} F_{\tau}(s,k;\mathcal R_{1b}) =   \sum\limits_{j = s }^{\left\lfloor {(\tau - 2k + s - 2)/3} \right\rfloor } &{\left( {\begin{array}{*{20}c}
   {\tau - 2j - 2k - 1}  \\
   {j - s + 1}  \\
 \end{array} } \right)} \nonumber \\ &\times \frac{1}{\left|{\rm Det}({\bf M}_\tau-{\bf 1})\right|} \label{bottomlocal}\end{align}
and the sum over all boxes on the bottom row of $\mathcal R_1$ is
\beq 
_1F_{\tau b} =  \sum\limits_{s = 0}^{k
- 1} {\left( {\begin{array}{*{20}c}
   {k - 1}  \\
   s  \\
 \end{array} } \right)}  F_{\tau}(s,k;\mathcal R_{1b}). \eeq

The more complicated sums that arise for region $\mathcal R_1$ may be simplified for certain calculations of interest. In particular, consider the calculation of the variance for the Hannay-Ozorio sum in Sect.~\ref{sumfluct}. Equations \refeq{R1local} and \refeq{bottomlocal}, like their $\mathcal R_4$ counterpart Eq.~\refeq{hannaysum}, are sums of the form of Eq.~\refeq{appendixform} from Appendix~\ref{sect:sumformula} with $\alpha = 4$, and they may both be expressed as 
\beq 
F_{\tau}(s; \mathcal R_{1}) = 4^{-k} + A'e^{\gamma s} + \overline{A'e^{\gamma s}} 
\eeq 
where $A' = c_{2}2^{-\tau /2 - k +1/2}e^{i\theta (\tau - 2k + 1)}$ and $\gamma = \ln2 - 2\theta i$.

For the sum of squared deviations over boxes in $\mathcal R_1$  the expression to evaluate is
\begin{align}
\sigma^2(\tilde F_{\tau}) = & \hspace{1 mm} 4^{ - k}  \sum\limits_{v = 1}^{k - 1} {\sum\limits_{t = 0}^{k - 1} {\left( {\begin{array}{*{20}c}
   {k - 1}  \\
   v  \\
\end{array}} \right)} \left( {\begin{array}{*{20}c}
   {k - 1}  \\
   t  \\
\end{array}} \right)\tilde{F}_{\tau}(t+v)^2 \quad } \nonumber \\
  &+ 4^{-k}\sum\limits_{s = 0}^{k - 1} {\left( {\begin{array}{*{20}c}
   {k - 1}  \\
   s  \\
\end{array}} \right)\tilde{F}_{\tau}(s)^2 }  \quad
\end{align}
where $\tilde{F}_{\tau}(s) = F_{\tau}(s) - 4^{-k}$.

There is an identity due to Vandermonde~\cite{Graham94} which simplifies the double sum above and gives a result that is almost exactly like the sum for region $\mathcal R_4$.  Denoting the double sum by $W$ gives
\begin{align}
W &= \sum\limits_{v = 1}^{k - 1} {\sum\limits_{t = 0}^{k - 1} {\left( {\begin{array}{*{20}c}
   {k - 1}  \\
   v  \\
\end{array}} \right)} \left( {\begin{array}{*{20}c}
   {k - 1}  \\
   t  \\
\end{array}} \right)\tilde{F}_{\tau}(t+v)^2 }\quad \nonumber \\
 &=  \sum\limits_{v = 1}^{k - 1} {\sum\limits_{s = v}^{v + k - 1} {\left( {\begin{array}{*{20}c}
   {k - 1}  \\
   v  \\
\end{array}} \right)\left( {\begin{array}{*{20}c}
   {k - 1}  \\
   {s - v}  \\
\end{array}} \right)\tilde{F}_{\tau}(s)^2 } }
\end{align}
where $s = t + v$. Interchanging the order of summation and breaking this into two sums generates
\begin{align}
W = &\sum\limits_{s = 1}^{k - 1} {\tilde{F}_{\tau}(s)^2 \sum\limits_{v = 1}^s {\left( {\begin{array}{*{20}c}
   {k - 1}  \\
   v  \\
\end{array}} \right)\left( {\begin{array}{*{20}c}
   {k - 1}  \\
   {s - v}  \\
\end{array}} \right)} } \nonumber \\
 &+ \sum\limits_{s = k}^{2k - 2} {\tilde{F}_{\tau}(s)^2 \sum\limits_{v = s - k + 1}^{k - 1} {\left( {\begin{array}{*{20}c}
   {k - 1}  \\
   v  \\
\end{array}} \right)\left( {\begin{array}{*{20}c}
   {k - 1}  \\
   {s - v}  \\
\end{array}} \right)} }. \,
\nonumber \end{align}
Vandermonde's convolution identity is
\beq\sum\limits_b {\left( {\begin{array}{*{20}c}
   a  \\
   b  \\
\end{array}} \right)} \left( {\begin{array}{*{20}c}
   c  \\
   {d - b}  \\
\end{array}} \right) = \left( {\begin{array}{*{20}c}
   {a + c}  \\
   d  \\
\end{array}} \right)\eeq
where the sum is over all values of $b$ for which the summand is not zero. This gives
\[
\sum\limits_{v = 1}^s {\left( {\begin{array}{*{20}c}
   {k - 1}  \\
   v  \\
\end{array}} \right)\left( {\begin{array}{*{20}c}
   {k - 1}  \\
   {s - v}  \\
\end{array}} \right)} \,\,\,\, = \,\,\,\left( {\begin{array}{*{20}c}
   {2k - 2}  \\
   s  \\
\end{array}} \right) - \left( {\begin{array}{*{20}c}
   {k - 1}  \\
   s  \\
\end{array}} \right)
\]
when $1 \leq s \leq k-1$, where the subtracted term corresponds to $v = 0$.  Also note that
\[
\sum\limits_{v = s - k + 1}^{k - 1} {\left( {\begin{array}{*{20}c}
   {k - 1}  \\
   v  \\
\end{array}} \right)\left( {\begin{array}{*{20}c}
   {k - 1}  \\
   {s - v}  \\
\end{array}} \right)} \,\,\,\, = \,\,\,\left( {\begin{array}{*{20}c}
   {2k - 2}  \\
   s  \\
\end{array}} \right)
\]
when $k \leq s \leq 2k - 2$.  Combining the two gives
\[ W =
\sum\limits_{s = 1}^{2k - 2} {\left( {\begin{array}{*{20}c}
   {2k - 2}  \\
   s  \\
\end{array}} \right)\tilde{F}_{\tau}(s)^2 } \,\, - \sum\limits_{s = 1}^{k - 1} {\left( {\begin{array}{*{20}c}
   {k - 1}  \\
   s  \\
\end{array}} \right)\tilde{F}_{\tau}(s)^2 },
\]
and therefore,
\begin{align}
\sigma^2(\tilde F_{\tau}) = 4^{ - k} [ &\sum\limits_{s = 1}^{2k - 2} {\left( {\begin{array}{*{20}c}
   {2k - 2}  \\
   s  \\
\end{array}} \right)\tilde{F}_{\tau}(s)^2 } \,\, \nonumber \\ &-  \sum\limits_{s = 1}^{k - 1} {\left( {\begin{array}{*{20}c}
   {k - 1}  \\
   s  \\
\end{array}} \right)\tilde{F}_{\tau}(s)^2 } \,\, \nonumber \\ &+ \quad \sum\limits_{s = 0}^{k - 1} {\left( {\begin{array}{*{20}c}
   {k - 1}  \\
   s  \\
\end{array}} \right)\tilde{F}_{\tau}(s)^2 }]
\end{align}
or
\begin{align}
\sigma^2(\tilde F_{\tau}) &= 4^{ - k} [ {\sum\limits_{s = 1}^{2k - 2} {\left( {\begin{array}{*{20}c}
   {2k - 2}  \\
   s  \\
\end{array}} \right)\tilde{F}_{\tau}(s)^2 } \,\,\,\, + \,\,\,\tilde{F}_{\tau}(0)^2 } ] \nonumber \\
 &=  4^{ - k} \sum\limits_{s = 0}^{2k - 2} {\left( {\begin{array}{*{20}c}
   {2k - 2}  \\
   s  \\
\end{array}} \right)\tilde{F}_{\tau}(s)^2 } \,\quad
\end{align}
which is simply
\beq 
\sigma^2(\tilde F_{\tau}) =
4^{ - k} \sum\limits_{s = 0}^{2k - 2} {\left( {\begin{array}{*{20}c}
   {2k - 2}  \\
   s  \\
\end{array}} \right)(A'e^{\gamma s}  + \overline {A'^{\gamma s} } \,\,)^2 }. \quad
\eeq
This sum is of exactly the same form as Eq.~\refeq{variance2} for computing the variance for the other regions of the unit square.

\section{A sum formula for the SR map}
\label{sect:sumformula}  

Upon examination of the form of Eq.~\refeq{hannaysum} and given the result of Eq.~\refeq{inversedet}, it happens that in order to arrive at closed form expressions for fluctuations in the Hannay-Ozorio sum, Eq.~\refeq{HannayOzorio}, it turns out that several sums of the form 
\beq
S(n,\alpha) = \sum_{i=0}^{\lfloor n/3 \rfloor} {\left( {\begin{array}{*{20}c}
   {n - 2i}  \\
   i  \\
 \end{array} } \right)} \alpha ^i 
 \label{appendixform}
 \eeq 
 are needed for various real values of $\alpha$, with $n$ a positive integer.  This section gives a general discussion of such sums and presents a method of finding closed form expressions for them before moving on to the main results of interest. In the analysis of the SR map, the cases $\alpha$ = 1, 4, 16 and -16 show up naturally when considering the lower order moments of Sect.~\ref{moments}.

Obtaining a closed form for the sum may begin by finding a recursion formula for it, and then using a standard technique for solving such recursions.  Recall the recursion for building Pascal's triangle: $\left( {\begin{array}{*{20}c}
   k  \\
   i  \\
 \end{array} } \right) = \left( {\begin{array}{*{20}c}
   {k - 1}  \\
   {i - 1}  \\
\end{array} } \right) + \left( {\begin{array}{*{20}c}
   {k - 1}  \\
   i  \\
\end{array} } \right)$ where $k$ and $i$ are greater than one.  Applying this gives
\begin{align}
  S(n,\alpha) &= \sum_{i=0}^{\lfloor n/3 \rfloor} {\left( {\begin{array}{*{20}c}
   {n - 2i}  \\
   i  \\
 \end{array} } \right)} \alpha ^i \nonumber\\ &= \sum_{i=1}^{\lfloor n/3 \rfloor} {\left( {\begin{array}{*{20}c}
   {n - 2i - 1}  \\
   {i - 1}  \\
 \end{array} } \right)\alpha ^i } \nonumber\\ & \hspace{1 cm} + \sum_{i=0}^{\lfloor (n-1)/3 \rfloor} {\left( {\begin{array}{*{20}c}
   {n - 2i - 1}  \\
   i  \\
 \end{array} } \right)\alpha ^i } \nonumber\\
  &= \alpha \sum_{j=0}^{\lfloor (n-3)/3 \rfloor} {\left( {\begin{array}{*{20}c}
   {n - 3 - 2j}  \\
   j  \\
 \end{array} } \right)\alpha ^j} \nonumber \\
 & \hspace{1 cm} + \sum_{i=0}^{\lfloor (n-1)/3 \rfloor} {\left( {\begin{array}{*{20}c}
   {n - 1 - 2i}  \\
   i  \\
 \end{array} } \right)\alpha ^i }
 \end{align}
 where $j=i-1$ in the first summation.  This produces the recursion relation 
 \beq S(n,\alpha) = S(n-1,\alpha) + \alpha S(n-3,\alpha) 
 \label{eqn:recursion}
 \eeq 
for $n \geq 3$ with initial conditions  $S(0,\alpha) = S(1,\alpha) = S(2,\alpha) = 1$.  Of historical note, the 14$^{th}$ century  Indian mathematician Narayana studied a problem of the proliferation of cows (each offspring gives birth after its third year) that leads to this very same recursion relation with $\alpha=1$~\cite{Datta93}.  A standard technique for solving such recurrence relations is to
look for solutions of the form $S(n,\alpha) = \beta^n$, and thus
$S(n-1,\alpha) = \beta^{-1}\beta^{n}, S(n-3,\alpha) =
\beta^{-3}\beta^n$. Plugging these expressions into
Eq.~\refeq{eqn:recursion} the factor $\beta^n$ cancels and leaves the cubic equation 
\beq 
\beta ^3 - \beta ^2 - \alpha = 0.
\label{eqn:cubic}
\eeq 
The three roots of this cubic $\beta_1$,
$\beta_2$, $\beta_3$  give three solutions of the recurrence
$\beta_1^n$, $\beta_2^n$, and $\beta_3^n$ . The difference equation
Eq.~\refeq{eqn:recursion} is third order, linear, and homogeneous, and the
standard theory of such difference equations (analogous to that for differential equations) says that if there are three linearly independent solutions, then the general solution may be formed as a linear combination of these independent solutions. Naturally this cubic equation also appears when using the matrices from symbolic dynamics. Indeed the characteristic equations for $T_0$, $T_1$ and $T_2$ are, up to a scaling, the same as the cubic equations with $\alpha=1,4$ and 16, respectively.

The cubic polynomial $f(\beta) = \beta^3 - \beta^2 - \alpha$ has a local maximum of $-\alpha$ when $\beta = 0$ and a local minimum at $\beta = 2/3$.  It has one real root, say $\beta_1$, when $\alpha > 0$ and also when $\alpha < -4/27$, which covers all of the cases of interest. The other two roots are complex conjugates, $\beta_2 = re^{i\theta} = \delta + i\gamma$ and $\beta_3 = re^{-i\theta} = \delta - i\gamma$. Since $\beta_{1}^2(\beta_1 - 1) = \alpha$, it implies that $\beta_1 > 1$ when $\alpha >0$, and $\beta_1 < -1/3$ when $\alpha < -4/27$, so that $\beta_1$ has the same sign as $\alpha$.

Because the three roots are distinct, the three solutions are independent and the general solution to Eq.~\refeq{eqn:recursion} may be written as 
\beq S(n,\alpha)  = c_1 \beta_1^n \, + \,c_2 (re^{i\theta } )^n \, + \,\,c_3 (re^{ - i\theta } )^n \label{sum1}
\eeq 
where the coefficients $c_1$, $c_2$, $c_3$ may be found from the initial conditions on the
recurrence (in all cases here the initial conditions on $S(n,\alpha)$ are real).  

It is possible to express the two complex roots, as well as the coefficients, in terms of the real root $\beta_1$ and in terms of $\alpha$.  The constant $c_1$ is real and $c_3$ is the conjugate of $c_2$, which can be denoted as $c_2$ = $a$ + $i b$, $c_3$ = $a$ - $i b$ where $a$ and $b$ are real.  The polynomial of Eq.~\refeq{eqn:cubic} may be factored as $(\beta -  \beta_1)(\beta - \delta - i\gamma)(\beta - \delta + i\gamma)$.

Comparing the constant terms of the polynomial written both ways gives $\beta_1(\delta^2 + \gamma^2) = \alpha$.  Thus, the magnitude of the complex roots is $r = (\delta^2 + \gamma^2)^{1/2} = (\alpha / \beta_1)^{1/2}$.   Because $\beta_{1}^2 - \beta_1 = \alpha/\beta_1$, it is also true that $r < \beta_1$ when $\alpha > 0$, but $r > \beta_1$ when $\alpha < -4/27$. The significance is that for large $n$ the oscillatory terms in Eq.~\refeq{sum1} are dominated by the first term when $\alpha > 0$, but the oscillatory terms are dominant when $\alpha < -4/27$. This observation is used ahead in deriving several asymptotic results.

Comparing the coefficients of the square terms gives $\delta = (1 - \beta_1)/2$, which is a negative number when $\alpha > 0$ and a positive number when $\alpha < -4/27$.  The two complex roots are in the second and third quadrants when $\alpha > 0$, and in the first and fourth quadrants in the other case. For $\alpha > 0$, we can take $\theta = \pi + \arcsin((1 - \beta_1)/2r)$ which lies in the second quadrant, and for $\alpha < -4/27$ we can take $\theta =  \arcsin((1 - \beta_1)/2r)$ which is in the first quadrant.

With the complex roots in terms of the one real root, it suffices to find the real root, which can be expressed straightforwardly for the regime of interest, i.e.~either $\alpha>0$ or $\alpha<-4/27$.  In that case, with $x=1+27\alpha/2$,
\begin{equation}
\label{realroot}
\beta_1 = \frac{1}{3}\left[ 1+ \left(x+\sqrt{x^2-1}  \right)^{1/3} + \left(x - \sqrt{x^2-1} \right)^{1/3} \right].
\end{equation}
Rewriting Eq.~\refeq{sum1} in terms of the real constants $c_1$, $a$, $b$ gives
 \beq 
 S(n,\alpha)  = c_1 \beta _1^n \, + 2 \left(\frac{\alpha}{\beta _1}\right)^{n/2}\left[ a\cos(n\theta) - b\sin(n\theta) \right]
 \label{genericform} 
 \eeq
Putting in the initial conditions gives three real equations for the coefficients which may be solved in terms of $\alpha$ and $\beta_1$.  Skipping the algebraic steps, one finds
\begin{align} c_1  &= \frac{{\alpha  + \beta _1^2 }} {{3\alpha  + \beta _1^2
}},\,\,\,a = \frac{\alpha } {{3\alpha  + \beta _1^2 }},\,\,\, \nonumber \\
b &= - \frac{{\beta _1 }} {{(3\alpha  + \beta _1^2 )}}\left(\frac{\alpha}{3\beta
_1  + 1}\right)^{1/2},  
\end{align}
which gives a complete and explicit solution to Eq.~\refeq{appendixform}.

The counting results for periodic points in the previous section beg the question, `what is the asymptotic density of the combinatorial expression
 $\left( {\begin{array}{*{20}c}
    {n - 2i}  \\
    i  \\
 \end{array} } \right)$ as a function of $i$?'.  It turns out that it is possible to find the asymptotic mean, variance, and density of $i$ (as $n$ approaches infinity) by essentially the same algebraic methods used in the recurrence relation.  Using the moment-generating function, or by using Stirling's approximation (essentially a saddle point expression), it can be shown that the density of $\left( {\begin{array}{*{20}c}
   {n - 2i}  \\
   i  \\
 \end{array} } \right)$, when properly normalized, converges to a normal density. The moment-generating function technique also gives simple formulas for the asymptotic mean and variance.  More specifically, 
\beq 
S(n,e^t) = \sum_{i=0}^{\lfloor n/3 \rfloor} {\left( {\begin{array}{*{20}c}
     {n - 2i}  \\
     i  \\
\end{array} } \right)} e^{ti} 
\eeq 
where one substitutes $\alpha = e^t$, and the moment-generating function for this density is $\phi(n,t) = S(n,e^t)/S(n,0)$.  Large $n$ gives $S(n,e^t) \rightarrow c(t)\beta_1(t)^n$ and  $\phi(n,t) = c(t)\beta_1(t)^n\left[c(0)\beta_1(0)^n\right]^{-1}$ where $\beta_1(t)$ is the real root of $\beta(t)^3 - \beta(t)^2 - e^t = 0$.   The details are omitted, but by differentiating the cubic equation all of the derivatives of $\phi(n,t)$ can be found, which can be used to find the moments of the density.  When $\alpha = e^t = 1$ the real root of Eq.~\refeq{eqn:cubic} is $\beta_1(0) = 1.46557123\ldots$, which in the next section is seen to have special significance to this map.  The asymptotic mean of this density can be shown to be 
 \begin{equation}
 \mu (n) =  \frac{1}{3+\beta^2_1(0)} \left[ n-2 +  \frac{6}{3+\beta^2_1(0)}\right]  
 \label{mean} 
 \end{equation}
 and the variance 
  \begin{equation} 
 \sigma^2 (n) = \frac{\beta^5_1(0)}{\left[3+\beta^2_1(0)\right]^3}\left[n - 2 +  \frac{12}{3+\beta^2_1(0)} \right] 
 \end{equation}
The scaling of the mean is $n$ and the width is $n^{1/2}$.  All the higher reduced cumulants (rescaled by the appropriate power of the width) vanish in the limit of $n\rightarrow \infty$.

\bibliography{rmtmodify,quantumchaos,classicalchaos,nano}

\end{document}

!!!!!!!!!!!!!!!!!!!!!!!!!!!!!!!!!!!!!!!!!!!!!!!!!!!!!!!!!!!!!!!!!!!!!!!!!!!!!!!!!!!!!!!!!!!!!!!!!!!!!!!!!!!!!!!!!!!!!!!!!!!!!!!!!!!!!!!!!!!!!!!!!!!!!!!!!!

!!!!!!!!!!!!!!!!!!!!!!!!!!!!!!!!!!!!!!!!!!!!!!!!!!!!!!!!!!!!!!!!!!!!!!!!!!!!!!!!!!!!!!!!!!!!!!!!!!!!!!!!!!!!!!!!!!!!!!!!!!!!!!!!!!!!!!!!!!!!!!!!!!!!!!!!!!

In semiclassical theory, quantum mechanical quantities are expressed in terms of functions, all of whose input information comes from the quantum system's classical analog.  Further analysis and manipulation of these functions to address quantum fluctuation properties often leads to results depending at some point on a classical sum rule.

In fact, exponential convergence implies a very short time scale in semiclassical dynamics which is logarithmic in $\hbar$; i.e $\tau_b=\ln (S_0/\hbar)/\lambda$ where $\lambda$ is the exponential convergence scale and $S_0$ is some characteristic classical action, perhaps something close to the action of the shortest periodic orbit.  As Berry noted in his semiclassical theory of spectral rigidity~\cite{Berry85}, this short time scale implies an outer energy scale inversely proportional to $\tau_b$ beyond which the spectral rigidity characteristic of random matrix theory, as asserted in the Bohigas-Giannoni-Schmit conjecture~\cite{Bohigas84}, does not apply.   Thus, the typical, ever-growing, stability matrix fluctuations cannot have any consequences for the two-point density of states fluctuations on an energy range shorter than the outer energy scale.

It does not necessarily follow from the above arguments that all possible quantum statistical properties of chaotic systems are void of information about classical sum rule fluctuations other than beyond a scale implied by a convergence exponent.  

Nevertheless, perhaps it is worthwhile to investigate more precisely quantum statistical properties in order to identify the mechanisms by which classical sum rule fluctuations vanish from the expressions.  Although, it would appear to be at odds with the Bohigas-Giannoni-Schmit conjecture~\cite{Bohigas84}, new classes of quantum statistical measures, possibly involving higher order correlations, transport fluctuations, and/or involving the eigenstates, may suggest ways in which the exponential convergence rate can be slowed down.  It is known that just considering the higher-order (greater than two-point) Dyson-Mehta cluster functions does not lead to sum rules substantially different than the Hannay-Ozorio sum rule~\cite{Shukla97}.  On the other hand, it turns out that the scaling behavior with particle number of the spatially and energy integrated statistical quantity that determines certain fluctuations of the many-electron ground state in a closed or nearly closed quantum dot is fundamentally altered by the presence of the bouncing ball modes in the stadium billiard~\cite{Ullmo09}.  Even for large enough particle number, where the fraction of bouncing ball modes is negligible, this statement is not altered.

The quantum bakers map has been studied widely over the years, and 
much is known about it. However the quantum lazy baker maps have been less studied except in the original work \cite{Lakshminarayan93}, nevertheless their quantization is once again simple and provides us a possibility to observe the fluctuations discussed herein in appropriate quantum mechanical measures.

In \refsect{sect:sumformula} we give a discussion of a general mathematical problem which is of critical importance in arriving at analytic results for the problem at hand.


\section{Statistics and Results, Old}

 \label{sec:statistics}
 Statistical properties of the quantity $\frac{1}{|det(M-I)|}$
are of interest for several reasons [4]; this quantity is sensitive
to variations over the periodic points and thus provides a
convenient measure of the fluctuations of the map. It is also a
quantity of fundamental importance in semiclassical mechanics. The
finite-time fluctuations over phase space, and with increased time,
of this and related quantities are investigated.

\subsection{Fluctuations in Uniformity}
I first look at the quantity \beq \sum_{f.p.\in{A}}
{\frac{1}{|det(M-I)|}} \label{detj}
 \eeq at fixed period, where the sum is over fixed points in a
 box of area A, and I will initially take A
 to be the whole space $[0,1] \times [0,1]$. The theoretical prediction of the uniformity principle
 is that this quantity should come out to be equal to the area of the box, A.
Computing this value starting at period $T$ = 5 it is seen that this
quantity rapidly converges to 1, i.e. the uniformity principle is
quickly recovered.
 $\sum{\frac{1}{|det(M-I)|}}$ computed over the whole space and labeled $total$ is
 shown against time
 in Table \ref{tab:totalhT} and \reffig{fig:totalhT}. The rapid convergence to uniformity is apparent. By
period 15 the agreement is within two decimal places and by
period 27 the agreement is within four decimal places of unity. \\
\begin{table}[!h]
\begin{center}
\caption{$total$ and $h(T)$}
\begin{tabular}{|c|c|c||c|c|c|}
   \hline
 T & total & h(T) & T & total & h(T) \\
  \hline
  5 & 0.493827 & 0.366204 & 18 & 0.994546 & 0.382207 \\
  6 & 1.665513 & 0.377008 & 19 & 1.002101 & 0.382222 \\
  7 & 1.348550 & 0.374467 & 20 & 1.002428 & 0.382300 \\
  8 & 0.974517 & 0.377911 & 21 & 0.999418 & 0.382235 \\
  9 & 0.964899 & 0.380666 & 22 & 1.000147 & 0.382234 \\
  10 & 0.995517 & 0.380878 & 23 & 1.000316 & 0.382241 \\
  11 & 0.929754 & 0.381226 & 24 & 0.999383 & 0.382242 \\
  12 & 0.996298 & 0.381757 & 25 & 0.999792 & 0.382243 \\
  13 & 1.041154 & 0.381936 & 26 & 1.000301 & 0.382244 \\
  14 & 1.009273 & 0.382007 & 27 & 1.000024 & 0.382244 \\
  15 & 0.994215 & 0.382105 & 28 & 0.999946 & 0.382245 \\
  16 & 1.003145 & 0.382163 & 29 & 1.000071 & 0.382245 \\
  17 & 0.996312 & 0.382186 & 30 & 0.999988 & 0.382245 \\
  \hline
\end{tabular} \\
\label{tab:totalhT}
\end{center}
\end{table}

One may compare the rate of convergence of
$\sum{\frac{1}{|det(M-I)|}}$ over the whole space to that of the
topological entropy,
  given by \beq
h(T) = \frac{1}{T} ln(p(T)) \label{$hT$} \eeq where $p(T)$ is the
number of fixed points of the period $T$ map. $h(T)$ is also
included in Table \ref{tab:totalhT} as well as a plotted against
time in \reffig{fig:totalhT}.
\begin{figure}
(a)\includegraphics[width=.4\textwidth]{figs/total.eps}
(b)\includegraphics[width=.4\textwidth]{figs/h(T).eps}
  \caption{$total$ and $h(T)$ against period.
  (a) {$total$} (b) ${h(T)}$  }
  \label{fig:totalhT}
 \end{figure}
 $h(T)$ quickly
converges to ~0.38 by approximately period 10, and by period 15
$h(T)$ has come within ~0.0001 of its value at $T$ = 30. Solving the
difference equation \refeq{recursion} below for $p(T)$ one finds for
the topological entropy a value very close to what is computed here
[2].
 In both the case of
$h(T)$ and of $total$, the quantities start to converge closely to
their large-time limits around periods 10 to 15. The rate of this
convergence for both quantities is determined by the sample size of
fixed points in the space. Using the recursion relation  \beq p(T) =
p(T-1) + p(T-3) \label{recursion}\eeq  $p(T)$ is shown against $T$
in Table \ref{tab:p(T)} out to period 40. Between periods 10 and 15
the number of fixed points jumps from 45 to 308, providing reliable
values for both $h(T)$ and $\sum{\frac{1}{|det(M-I)|}}$.
\\
\begin{table}
\begin{center}
\caption{p(T)}
\begin{tabular}{|c|c||c|c||c|c||c|c|}
  \hline
 T & p(T) & T & p(T) & T & p(T) & T & p(T) \\
  \hline
  5 & 5 & 14 & 210 & 23 & 6578 & 32 & 205220 \\
  6 & 9 & 15 & 308 & 24 & 9641 & 33 & 300765 \\
  7 & 14 & 16 & 452 & 25 & 14130 & 34 & 440793 \\
  8 & 20 & 17 & 663 & 26 & 20709 & 35 & 646014 \\
  9 & 30 & 18 & 972 & 27 & 30351 & 36 & 946780 \\
  10 & 45 & 19 & 1452 & 28 & 44482 & 37 & 1387574 \\
  11 & 66 & 20 & 2089 & 29 & 65192 & 38 & 2033589 \\
  12 & 97 & 21 & 3062 & 30 & 95544 & 39 & 2980370 \\
  13 & 143 & 22 & 4488 & 31 & 140027 & 40 & 4367945 \\
  \hline
\end{tabular} \\
\label{tab:p(T)}
\end{center}
\end{table}

In order to look at fluctuations from uniformity it is necessary to
divide the phase space into small boxes and look at the difference
from uniformity for each box at fixed time. I calculate here the
mean and variance between boxes of the quantity $diff$ and make a
density plot of the absolute value of $diff$ in order to see which
regions fluctuate the most from uniformity ($diff$ is the difference
between $\sum{\frac{1}{|det(M-I)|}}$ and the area of the given box).
I then increase $T$ and repeat and look at time variations in these
quantities. As $T$ is increased the number of periodic points starts
to grow exponentially ( for large $T$, $p(T)\cong 1.47^{T}$ ). This
being so one may shrink the box size and add more boxes so as to
gain insight into the finer structure of the fluctuations. This is
apparent in the density plots.

 The choice of how to shrink the box size with time
can be done in more than one way, and after considerable trial and
error I have chosen the following approximate algorithm for choosing
the box size as a function of time; keep the smallest number of
fixed points per box in the range $\sim$ 10 - 15 to avoid a blow up
in quantities calculated for empty boxes, and at the same time make
the average number of fixed points per box as small as possible
given the first criterion. This provides a trade off between finer
structure and continuing to have statistically significant values.
An additional justification of this choice is the following; we know
that for large $T$ the number of fixed points grows as $p(T)\sim
1.47^{T}$. It seems reasonable that an estimate for the box length,
$L_{est}$, should go as $L_{est}\sim \frac{1}{\sqrt{p(T)}}$ which
gives $\frac{1}{L_{est}} \sim 1.47^{T/2}$ or $L_{est}\sim
\frac{1}{(\sqrt{1.47})^{T}}$. My choice of the quantity $boxsize$,
the length of a box, and the value of $L_{est}$ are plotted together
in \reffig{fig:boxsize} for the largest computed values of $T$. The
similar trend with increased $T$ is apparent.

\begin{figure}[!h]
\includegraphics[width=.5\textwidth]{figs/boxsize.eps}
  \caption{choice of $boxsize$ and theoretical quantity $L_{est}$
  against period
   }
   \label{fig:boxsize}
 \end{figure}

 The values $meandiff$, the average of $diff$ between boxes,
and $vardiff$, the variance of $diff$ between boxes, are given for
values of $T$ between 20 and 37 in Table \ref{tab:meandiff} along
with the box length chosen for that $T$. Density plots of the value
of $|diff|$ for each box are shown for several $T$ in
\reffig{fig:difffig}. One can see from the table that the mean and
variance between the boxes of $diff$ look to be heading towards 0 as
$T$ increases, as we would expect. The density plots, whose largest
values are colored dark red and whose smallest values are colored
dark blue, show interesting information about the fluctuations from
uniformity, particularly at large periods. As with the periodic
points themselves, the regions of largest fluctuation are around the
period 3 orbit and the finer structure becomes apparent with
increased $T$.

\begin{figure}
(a)\includegraphics[width=.4\textwidth]{figs/difffigT35.eps}
(b)\includegraphics[width=.4\textwidth]{figs/difffigT37.eps}
  \caption{Density plot of the quantity $|diff|$ at periods
   (a) {T = 35} (b) {T = 37} }
 \label{fig:difffig}
 \end{figure}

\begin{table}
\begin{center}
\caption{meandiff, vardiff, and boxsize}
\begin{tabular}{|c|c|c|c||c|c|c|c|}
  \hline
 T & boxsize & meandiff & vardiff & T & boxsize & meandiff & vardiff \\
  \hline
  20 & 1/7 & 4.954906e-5 & 8.141150e-6 & 29 & 1/17 & 2.447304e-7 & 2.643265e-8 \\
  21 & 1/9 & 8.474588e-4 & 4.001953e-6 & 30 & 1/19 & -3.212057e-8 & 7.458974e-9 \\
  22 & 1/9 & 1.812514e-6 & 6.230262e-6 & 31 & 1/22 & -1.183830e-7 & 1.153524e-8 \\
  23 & 1/10 & 3.164661e-6 & 5.060538e-7 & 32 & 1/24 & 2.390146e-8 & 2.603853e-8 \\
  24 & 1/10 & 2.314075e-5 & 3.355925e-7 & 33 & 1/27 & 6.046294e-6 & 4.136039e-9 \\
  25 & 1/12 & -1.445257e-6 & 7.996814e-7 & 34 & 1/30 & -1.082679e-8 & 2.804168e-9 \\
  26 & 1/13 & 1.783413e-6 & 7.983111e-8 & 35 & 1/32 & 2.710499e-9 & 4.766274e-11 \\
  27 & 1/14 & 1.205265e-7 & 1.267462e-7 & 36 & 1/37 & 2.051693e-7 & 2.815803e-10\\
  28 & 1/16 & -2.127099e-7 & 2.231091e-9 & 37 & 1/42 & -2.994942e-9 & 6.392117e-10\\
  \hline
\end{tabular} \\
\label{tab:meandiff}
\end{center}
\end{table}

\subsection{Fluctuations in $\frac{1}{|det(M-I)|}$}
Apart form the uniformity principle it is also of considerable
interest to look at properties of $\frac{1}{|det(M-I)|}$ over
periodic points without summing, most notably because of the
potential of comparison of the simulation to analytic results (see
Section \ref{sec:analytics}).
 I have calculated
the mean and variance of this quantity within each box, which I will
call $meanwithin$ and $varwithin$, for a wide range of values of $T$
and have provided density plots of these quantities at several
different times. For consistency I have taken the choice of box size
to be the same as in the last analysis. Periods $T$ = 28 and $T$ =
35 are of the most interest since the box sizes chosen for these
times are binary; 1/16 and 1/32 respectively. The analytic method
for calculating the mean and variance of $\frac{1}{|det(M-I)|}$ uses
binary boxes so these will be the most important for comparison.
 I have also calculated the mean of $varwithin$ between boxes and the
variance of $varwithin$ between boxes, denoted $meanvarwithin$ and
$varvarwithin$, and show these quantities in Table
\ref{tab:meanvarwithin} at different times. These quantities appear
to be tending to 0 with time. Again, the density plots shown in
\reffigs{fig:meanwithin}{fig:varwithin} show an interesting
structure resemblant of the periodic points and which highlight
regions of increased stability. The data used to make these figures
have been saved and should provide an interesting comparison with
analytic results when these are completed.
\begin{figure}[!h]
(a)\includegraphics[width=.4\textwidth]{figs/meanwithinfigT35.eps}
(b)\includegraphics[width=.4\textwidth]{figs/meanwithinfigT37.eps}
  \caption{Density plot of $meanwithin$ at periods
   (a) {T = 35} (b) {T = 37} }
  \label{fig:meanwithin}
 \end{figure}

\begin{figure}[!h]
(a)\includegraphics[width=.4\textwidth]{figs/varwithinfigT35.eps}
(b)\includegraphics[width=.4\textwidth]{figs/varwithinfigT37.eps}
  \caption{Density plot of $varwithin$ at periods
   (a) {T = 35} (b) {T = 37} }
  \label{fig:varwithin}
 \end{figure}

\begin{table}[!h]
\begin{center}
\caption{meanvarwithin, varvarwithin, and boxsize}
\begin{tabular}{|c|c|c|c||c|c|c|c|}
  \hline
 T & boxsize & meanvarwithin & varvarwithin & T & boxsize & meanvarwithin & varvarwithin\\
  \hline
  20 & 1/7 & 1.711328e-6 & 4.218219e-12 & 29 & 1/17 & 7.882980e-9 & 2.355206e-16 \\
  21 & 1/9 & 1.604750e-6 & 6.843775e-12 & 30 & 1/19 & 5.311445e-9 & 1.969280e-16 \\
  22 & 1/9 & 5.466083e-7 & 4.674456e-13 & 31 & 1/22 & 3.059294e-9 & 6.445036e-17 \\
  23 & 1/10 & 3.185711e-7 & 1.918752e-13 & 32 & 1/24 & 1.263252e-9 & 7.352292e-18 \\
  24 & 1/10 & 1.652076e-7 & 9.475197e-14 & 33 & 1/27 & 7.471162e-10 & 4.668083e-18 \\
  25 & 1/12 & 1.095132e-7 & 2.965186e-14 & 34 & 1/30 & 3.451796e-10 & 6.880293e-19 \\
  26 & 1/13 & 4.618048e-8 & 4.710698e-15 & 35 & 1/32 & 1.989696e-10 & 3.762713e-19 \\
  27 & 1/14 & 2.953116e-8 & 4.956302e-15 & 36 & 1/37 & 1.193904e-10 & 1.294464e-19\\
  28 & 1/16 & 1.547095e-8 & 1.464950e-15 & 37 & 1/42 & 6.152091e-11 & 3.077761e-20\\
  \hline
\end{tabular} \\
\label{tab:meanvarwithin}
\end{center}
\end{table}

\section{Analytic Results and Potential}
\label{sec:analytics}
 In contrast to the method of numerically computing the mean and
 variance of $\frac{1}{|det(M-I)|}$ over all periodic points falling within certain boxes, the
 possibility of predicting the distribution of this quantity based
 on the known symbolic dynamics and already calculated
 $\gamma_{m}$'s would provide a nice theoretical result and a great
 comparison with the former method. At the present time this problem
 has not been completed, however I believe I have made significant
 progress in working towards this goal with a few preliminary
 results and predictions as well as an idea of how to continue
 proceeding in this direction. I first give a derivation of the
 combinatorial formula, Equation \ref{eqn:totalcombo}, which gives a
 nice starting point for the reasoning of how to predict the
 distribution of $\frac{1}{|det(M-I)|}$. I then outline the progress
 I have made towards this answer as well as give a first prediction
 which checks successfully with the numerical results.
\subsection{Combinatorial Formula}
Here I give an argument for finding the expression in Equation
\ref{eqn:totalcombo} from [2] for $p(T)$, the number of periodic
points of period $T$, excluding the period 2 orbit on the boundary.
I also explicitly put in expressions for the upper limits on the
summations which are not in the original paper.

Consider first only points in $\mathcal R_4$. Let us fix $T$ and fix the index
$j$; this also fixes $m$, the length of the string $\gamma_{m}$. How
many different $\gamma_{m}$'s are possible for a given string length
with $j$ changes between 0 and 1, given that the string must start
with a 1 since we consider first only points in $\mathcal R_4$?

There are $m$-1 places to put the $j$ changes, so the number of ways
to do this is simply $\VectorII{m-1}{j}$. Solving Equation
\ref{eqn:period} for $m$ in terms of $T$ and $j$ we have for the
number of period $T$ points in $\mathcal R_4$ with $j$ changes the expression
$\VectorII{T-2j-3}{j}$. Letting $j$ run up to its maximum, the total
number of period $T$ points in $\mathcal R_4$ is \beq \sum_{j=0}^{\lfloor
T/3\rfloor -1}{\VectorII{T-2j-3}{j}}. \eeq The brackets around $T/3$
indicate the floor function of this quantity. Since points in $\mathcal R_4$ map
uniquely to $\mathcal R_3$ and $\mathcal R_2$ in the following two time steps, the number of
points in $\mathcal R_4$, $\mathcal R_3$, and $\mathcal R_2$ is 3 times the above quantity.

Now consider $\mathcal R_1$. One observes that for a $\gamma_{m}$ with fixed $j$
that after $T$ iterations $\mathcal R_4$ has been visited $j+1$ times. Since the
same must be true for $\mathcal R_3$ and $\mathcal R_2$, $\mathcal R_1$ will therefore have been visited
a total of $T-3j-3$ times. So for a given $T$ and $j$ there are
\beq\frac{1}{j+1} \VectorII{T-2j-3}{j}\eeq period $T$ orbits and for
each orbit $\mathcal R_1$ is visited $T-3j-3$ times, thus the total number of
period $T$ points in $\mathcal R_1$ is \beq\sum_{j=0}^{\lfloor T/3\rfloor
-1}{\frac{T-3j-3}{j+1} \VectorII{T-2j-3}{j}}.\eeq With some algebra
the summand becomes \beq \frac{T-3j-3}{j+1}
\frac{(T-2j-3)!}{j!(T-2j-3-j)!} =
\frac{(T-2j-3)!}{(j+1)!}\frac{T-3j-3}{(T-3j-3)!} =
\frac{(T-2j-3)!}{(j+1)!(T-3j-4)!} = \VectorII{T-2j-3}{j+1}. \eeq
Making the substitution $k=j+1$ and summing we have
\beq\sum_{k=0}^{\lfloor T/3\rfloor}{\VectorII{T-2k-1}{k}}. \eeq We
can add the single fixed point at the origin which did not originate
in $\mathcal R_4$ by letting the sum start at 0. Finally, switching the dummy
index back to $j$ for aesthetic purposes the number of period $T$
points in $\mathcal R_1$ becomes \beq\sum_{j=0}^{\lfloor
T/3\rfloor}{\VectorII{T-2j-1}{j}}. \eeq Combining this with the
result for $\mathcal R_4$, $\mathcal R_3$, and $\mathcal R_2$ we have the total number of period T
points, \beq \boxed{p(T) = 3\sum_{j=0}^{\lfloor T/3\rfloor
-1}{\VectorII{T-2j-3}{j}} + \sum_{j=0}^{\lfloor
T/3\rfloor}{\VectorII{T-2j-1}{j}}.} \eeq From this combinatorial
formula it is simply a matter of algebraic manipulation to show that
this formula satisfies the recursion relation, Equation
\ref{eqn:recursion}.

\subsection{Distribution of $\frac{1}{|det(M-I)|}$}
>From Section A we know that the total number of period $T$ points in
$\mathcal R_4$ is
 \beq p(T) = \sum_{j=0}^{\lfloor T/3\rfloor -1}{\VectorII{T-2j-3}{j}} \label{R4points} \eeq
 where $j$ is the number of changes of 0's and 1's. Let us restrict ourselves to a
 binary box in $\mathcal R_4$ of boundaries $\frac{a}{2^{k}}$,$\frac{b}{2^{k}}$ for
 some integers $a$,$b$, and $k$. This box has area $2^{-k}\times 2^{-k}$.
 The condition for a periodic point to be inside this box is dependent on
 the corner point $(\frac{a}{2^{k}},\frac{b}{2^{k}})$. Say a corner point has the
  binary representation $p = p_{1}p_{2}p_{3}...p_{k}$, $q =
  q_{1}q_{2}q_{3}...q_{k}$. Then every other point in the square
  shares the same first $k$ digits of its binary
  representation with the corner point. The number of 0 and 1 changes
  in the corner point then restricts the number available for all
  other points in the box. Any point in the box has the form $p = p_{1}p_{2}p_{3}...p_{k}www...$
$q = q_{1}q_{2}q_{3}...q_{k}www...$ where the $w$'s are some
combination of 0's and 1's. Since the points are periodic, the
bi-infinite string representing their state is at some time of the
form $p.q =
...wwwp_{k}...p_{3}p_{2}p_{1}.q_{1}q_{2}q_{3}...q_{k}www...wwwp_{k}...p_{3}p_{2}p_{1}$
since the string must repeat itself. Therefore the $\gamma_{m}$
string for any periodic point is of the form $\gamma_{m} =
q_{1}q_{2}q_{3}...q_{k}w...w_{i}p_{k}...p_{3}p_{2}p_{1}$ where $i =
m - 2k = T-2j-2k-2$ denotes the total number of $w$'s. Let $s_{q}$
and $s_{p}$ denote the total number of changes between 0 and 1 for
each of the corner points of a box and let $s = s_{q} + s_{p}$. Then
every periodic point in the box already has $s$ changes and this is
a lower bound on $j$ in the sum \refeq{R4points}. Depending on the
end points of the $w$'s there may be 1 or 2 additional changes apart
from those within the $w$'s themselves, this point still needs to be
figured out. If there are s transitions between 0 to 1 in the corner
points and $j$ total transitions in $\gamma_{m}$, there must be
$j-s$ transitions in the $i+1$ possible places in the $w$'s. So
there are $\VectorII{i+1}{j-s} = \VectorII{T-2j-2k-1}{j-s}$ ways to
do this. Determining the upper bound on $j$ for a period $T$ point
is a little tricky and can be special-cased for both $j$ and $s$
even or odd, but I believe in each case we have that $max(j) =
floor((T-2k+s-1)/3)$.
 So we have for the number of periodic points in a binary
box in $\mathcal R_4$ whose corner point has $s$ changes between 0 and 1 in both
p and q the expression \beq \label{eqn:maybe} p(T;box\in R4) =
\sum_{j=s}^{\lfloor (T-2k+s-1)/3\rfloor}{\VectorII{T-2j-2k-1}{j-s}}.
\eeq The same sum is valid for boxes in $\mathcal R_3$ and $\mathcal R_2$ since these are
merely rotations of $\mathcal R_4$. This suggests that all boxes in $\mathcal R_4$, $\mathcal R_3$, and
$\mathcal R_2$ whose corner points have the same value of $s$ should have the
same number of periodic points! This  conjecture is analyzed below.
The case for the number of periodic points in a box in $\mathcal R_1$ still
needs to be considered. Also, Equation \ref{eqn:maybe} needs to
checked and probably corrected as this derivation has been somewhat
heuristic. I believe the idea outlined here is however valid.

If we want to compute the average of a function such as
$\frac{1}{|det(M-I)|}$ this may now be done. From Equation
\ref{eqn:jacobian} we have that, depending on whether $j$ is even or
odd, \beq \frac{1}{|det(M-I)|} = \frac{1}{2^{m} + 2^{-m} \pm 2} \eeq
where m is the length of the string $\gamma_{m}$ but in either case
to very good approximation we have \beq \frac{1}{|det(M-I)|} \cong
2^{-m}. \eeq With the relation between $\gamma_{m}$ and $T$,  $T =
2j + 2 + m$, we then have \beq \label{eqn:approx1}
\frac{1}{|det(M-I)|} \cong 2^{-T+2j+2}. \eeq For fixed T and box
with $p(T;box\in R4)$ periodic points we then have \beq \left\langle
\frac{1}{|det(M-I)|} \right\rangle = \frac{1}{p(T;box\in
R4)}\sum_{j=s}^{\lfloor
(T-2k+s-1)/3\rfloor}{\VectorII{T-2j-2k-1}{j-s}}2^{-T+2j+2} \eeq and
\beq Var\left(\frac{1}{|det(M-I)|}\right) = \left\langle
\frac{1}{|det(M-I)|^{2}} \right\rangle - \left\langle
\frac{1}{|det(M-I)|} \right\rangle^{2} \eeq \beq
 = \frac{1}{p(T;box\in R4)}\sum_{j=s}^{\lfloor
(T-2k+s-1)/3\rfloor}{\VectorII{T-2j-2k-1}{j-s}}2^{-2T+4j+4} -
\left\langle \frac{1}{|det(M-I)|} \right\rangle^{2}. \eeq We are
then reduced to evaluating these combinatorial sums to find analytic
estimates for the mean and variance and potentially other moments of
the distribution. These sums may have known approximations or
possibly even closed form formulas, but even evaluating them
numerically would still be very fast compared to the method of
simulation over all periodic points and would make a great
comparison to that method. I plan to continue working towards this.

As mentioned above, even without completing the problem, one
analytic prediction that can be checked immediately with the
numerical results is the statement that all boxes which have the
same number of 0 and 1 changes for their corner points should
contain the same number of periodic points. To check this I produced
numerically the number of fixed points per box at $T$ = 28 and
$boxsize$ = 1/16 (binary choice, of course). I also produced the
binary string of the lower left corner point for each box. I do not
bother to add here the number of points in all 256 boxes or all 512
binary strings, but as an example there were 18 different boxes in
$\mathcal R_2$,$\mathcal R_3$, or $\mathcal R_4$ which each contained 19 fixed points. In Table
\ref{tab:gamma} I give the binary strings for the corner points of
each of these boxes. Indeed, all boxes containing exactly 19 fixed
points each have the same value for $s$, the total number of 0 and 1
changes for the corner point! This has been confirmed by many more
checks than just the one mentioned here. This hopefully provides
somewhat of a first check of the validity of the ideas outlined in
this section.

\begin{table}[!h]
\begin{center}
\caption{qc,pc, and s}
\begin{tabular}{|c|c|c||c|c|c|}
  \hline
\gamma_{-1} &=& h(T)
\end{eqnarray}
Just a few comments about this stuff.  There is a correction term to
the expression for $h(T)$ and it is well approximated by the
following
\begin{equation}
h(T) \approx \ln {1-2x_0 \over  1-3x_0} + {1\over 5} {\rm
e}^{-(1-2x_0)T}\cos (1.855T)
\end{equation}
I guess that I should know how to derive a correction of this kind,
but I haven't tried, just played around with fitting the correction.
Kind of curious.  Also, the uniformity principle implies that
$\gamma_{-1}=h(T)$ and that works!  Basically, plugging in $1/4$ for
$x_{-1}$, everything cancels except the $h(T)$ part.  Finally, the
fluctuations about the uniformity principle are governed by
$\gamma_{-2}$  ($\approx 0.351728124$).  So we should be in a
position to discuss analytically the fluctuations about the
uniformity principle now.

Cubic equations are a pain, but I think the solutions are in about
the simplest form possible.  Although, see the identity at the end.
I didn't expect that one.  Remember that $\gamma_n$ is supposed to
be related to an expansion over the cumulants of the probability
density for finite time $\lambda$.   As usual, the probability density
appears to go to a Gaussian (or log normal depending on the quantity
of interest), but too slowly for the expression to be truncated
after the variance term.  However, I did calculate (sent in an
e-mail earlier) an expression for the variance of the Gaussian
density.  It is
\begin{equation}
\sigma^2_x = \left[T\left( {9\over 1-3x_0} - {4\over 1-2x_0} +1
\right)\right]^{-1}
\end{equation}
and along with $x_0$, the mean, that fit the density numerically
quite well and it seemed to improve with increasing $T$ as it
should.

Did either of you know that
\begin{equation}
1=\left[ {2\over 7} + \left({3 \over 7}\right)^{3/2} \right]^{1/3} +
\left[ {2\over 7} - \left({3 \over 7}\right)^{3/2} \right]^{1/3}
\qquad ???
\end{equation}
To me anyway, this begs the question as to whether this is an
accidental relation or whether there is an infinity of identities of
the following structure:
\begin{equation}
c=\left[ a + b^{3/2} \right]^{1/3} + \left[ a - b^{3/2}
\right]^{1/3}
\end{equation}
where $\{a,b,c\}$ are all simultaneously rational numbers.  Calling
$\{a,b,c\}$ a solution, it is easy to see that $\{-a,b,-c\}$ is also
a solution.  More generally, it is easily verified that solutions
for all $c$ can be mapped onto solutions for $c=1$ as follows
\begin{equation}
\{a,b,c\} \longrightarrow \{{a\over c^3},{b\over c^2},1\}
\end{equation}
so that there is no loss of generality in considering only $c=1$
[the $c=0$ case is trivially $a=0$].  So the notation $\{a,b\}$ is a
solution for which $c=1$ and gives the most general class of
solutions by inverting the scaling given above.  I found one simple
way of proving that there is an infinity of solutions and generating
them.  Consider each term separately, one can always decompose
\begin{equation}
\left[ a \pm b^{3/2} \right]^{1/3} = {c\over 2} \pm \gamma
\end{equation}
where $\gamma$ is some unknown irrational number.  This can always
be done since adding any rational number to an irrational number
gives an irrational number.  This means that
\begin{equation}
a \pm b^{3/2}  = {c\over 2}^3 \pm 3 {c\over 2}^2 \gamma + 3 {c\over
2} \gamma^2 \pm \gamma^3
\end{equation}
or rather
\begin{eqnarray}
a &=& {c\over 2}\left( {c^2\over 4} + 3  \gamma^2\right)\\
b &=& \gamma^2 \left( {3c^2 \over  4\gamma^2} + 1\right)^{2/3}
\end{eqnarray}
>From the first relation, it is seen that $\gamma^2$ must be a
rational number, otherwise $a$ is not a rational number.  The logic
for enforcing $b$ to be a rational from the second relation, is only
slightly more complicated.  One cannot take the cube root of a
rational number and end up with the square root of a rational number
unless the cube root leaves a rational number to begin with.  So the
requirement to be enforced is that
\begin{equation}
\left( {3c^2 \over  4\gamma^2} + 1\right)^{1/3} = {r\over s}
\end{equation}
where $(r,s)$ are co-prime integers.  In addition, $r>s>0$ unless
one considers imaginary $\gamma$ (which I should do, but at this
point I have not done that yet nor considered complex rationals).
Inverting this relation generates,
\begin{equation}
\gamma^2 = {3\over 4}\cdot{ c^2 s^3 \over r^3-s^3}
\end{equation}
which is consistent with $\gamma^2$ being a rational number.
Therefore, choosing
\begin{eqnarray}
a &=& {c^3\over 8}\left( 1 + {9 s^3  \over r^3 -s^3} \right)\\
b &=& {c^2 \over 4}\cdot  {3sr^2 \over  r^3-s^3}
\end{eqnarray}
is guaranteed to generate a solution.  From this perspective, the
$(a,b) = (2/7,3/7)$ solution is just the $(r,s)=(2,1)$ solution.  We
can now generate an infinity of solutions for all $(r,s)$ co-prime
pairs.  The $(r,s)=(3,1)$ solution is $(a,b)=(35/208,27/104)$ and
the $(r,s)=(3,2)$ solution is $(a,b)=(91/152,27/38)$. The series of
solutions $(r,s)=(s+n,s)$ for $s$ increasing and $n$ fixed, gives an
infinite number of solutions converging to the point (i.e. slope)
$b/a=2/3$.

There is some reason to want to express the Diophantine relation as
an elliptic curve problem, which would mean a quadratic of a or b is
equated to a cubic of the other.  So far, the closest I could manage
is
\begin{equation}
b^3 = a^2 - \left( {1-2a \over 3} \right)^3
\end{equation}
which gives a cubic-cubic relation instead of a quadratic-cubic.  It
is possible to find other cubic-cubic relations, but up to the
point, I have not found a quadratic-cubic relation.

\end{document}